\documentclass[range]{ar2e}
\pdfoutput=1
\usepackage{lscape,graphicx}
\usepackage{rotating}
\usepackage{epstopdf}
\usepackage{color}
\usepackage{subfig}

\begin{document}




\title{Theoretical Perspectives on Protein Folding}

\markboth{Theoretical Perspectives on Protein Folding}{Thirumalai et. al.}

\author{D. Thirumalai,$^{1}$ Edward P. O'Brien,$^2$  Greg Morrison$^3$ and Changbong Hyeon$^4$
\affiliation{$^1$Biophysics Program, Institute for Physical Science and Technology and Department of Chemistry and Biochemistry, University of Maryland, College Park, MD 20742, USA; email: thirum@umd.edu\\
$^2$Department of Chemistry, University of Cambridge, Cambridge CB2 1EW, United Kingdom\\
$^3$School of Engineering and Applied Science, Harvard University, Cambridge, Massachusetts, 02138, USA\\
$^4$School of Computational Sciences, Korea Institute for Advanced Study, Seoul 130-722, Republic of Korea}}
\begin{abstract}
Understanding how monomeric proteins fold under {\it in vitro}
conditions is crucial to describing their functions in the cellular context. 
Significant advances both in theory and experiments
have resulted in  a conceptual framework for describing the folding
mechanisms of globular proteins. The experimental data and theoretical
methods have revealed the multifaceted character of proteins. Proteins exhibit universal features that can be determined using only
the number of amino acid residues ($N$) and polymer concepts. The
sizes of proteins in the denatured and folded states, cooperativity of
the folding transition, dispersions in the melting temperatures at the residue level, and time scales of folding  are to a large
extent determined by $N$.  The
consequences of finite $N$ especially on  how individual residues order upon
folding depends on the topology of the folded states.  Such intricate
details can be predicted using the Molecular Transfer Model that
combines simulations with measured transfer free energies of protein
building blocks from water to the desired concentration of the
denaturant.  By watching one molecule fold at a time, using single
molecule methods, the validity of the theoretically anticipated
heterogeneity in the folding routes,  and the $N$-dependent time scales
for the three stages in the  approach to the native state   have been established. Despite the
successes of theory, of which only a few examples are documented here, we
conclude that much remains to be done to solve the ``protein folding
problem" in the broadest sense.
\end{abstract}
\maketitle

\section{INTRODUCTION}
\begin{extract}
The quest to solve the protein folding problem in quantitative detail, which is surely only the first step in describing the functions of proteins in the cellular context, has led to great advances on both experimental and theoretical fronts \cite{Levitt75Nature,Matthews93ARBC,FershtBook,Ziv09PCCP,ThirumalaiACR96,Hyeon05BC,ThirumCOSB99,DobsonAngewante05,OnuchicARPC97,Dill08ARB,Shakhnovich06ChemRev,BaldwinTIBS99II,Schuler08COSB,EatonARBBS00,GruebeleNature03,BakerNature00,DillProtSci95,DillNSB97,Radford09NSMB,Clarke07COSB}.  
In the process our vision of the scope of the protein folding problem has greatly expanded. The determination of protein structures by X-ray crystallography  \cite{Kendrew61SciAm} and the demonstration that proteins can be reversibly folded following denaturation \cite{Anfinsen73Science} ushered in two research fields. The first is the prediction of the three dimensional structures given the amino acid sequence \cite{Baker05Science,Moult05COSB}, and the second is to describe the folding kinetics \cite{Schuler08COSB,OnuchicCOSB04,Hyeon05BC}. Another line of inquiry in the protein folding field opened with the discovery that certain proteins require molecular chaperones to reach the folded state \cite{Lorimer89Nature,SiglerNature97,Horwich06ChemRev,ThirumalaiARBBS01}.  More recently, the realization that protein misfolding, which is linked to a number of diseases, has provided additional wrinkles to the already complicated protein  folding problem \cite{ChitiARB06,SelkoeNature03,ThirumalaiCOSB03,Eisenberg06ACR}. Although known for a long time it is becoming more widely appreciated that the restrictions in the conformational space in the tight cellular compartments might have significant effect on  all biological processes including protein folding \cite{Cheung05PNAS,Minton08ARB}.  
In all these situations the protein folding problem is at the center stage. 
The solution to this problem requires a variety of experimental, theoretical and computational tools.  
Advances in all these fronts have given us hope that many aspects of, perhaps the ``simplest" of the protein folding problems, namely, how single domain globular proteins navigate the large dimensional potentially rugged free energy surface en route to the native structure are under theoretical control. 

Much of our understanding of the folding mechanisms comes from studies of proteins that are described using the two-state approximation in which only the unfolded and folded states are thought to be significantly populated.  However, proteins are  finite size branched polymers in which the native structure is only marginally stabilized by a number of relatively weak ($\sim O(k_BT)$) interactions.  From a microscopic point of view the unfolded state and even the folded state should be viewed as an ensemble of structures. Of course, under folding conditions the fluctuations in the native state are less than in the unfolded state.  In this picture rather than viewing protein folding as a unimolecular reaction ($U \leftrightarrow F$ where $U$ and $F$ being the unfolded and folded states respectively) 
one should think of the folding process as the interconversion of the conformations in the Denatured State Ensemble ({\bf DSE}) to the ensemble of structures in the Native Basin of Attraction ({\bf NBA}). The description of the folding process in terms of distribution functions  necessarily means that appropriate tools in statistical mechanics together with concepts in polymer physics \cite{deGennesbook,Florybook2,Edwardsbook, GrosbergBook} are required to understand the self-organization of proteins, and for that matter RNA \cite{Hyeon05BC}. 

Here, we provide theoretical perspectives on the thermodynamics and kinetics of protein folding of small single domain proteins with an eye towards understanding and anticipating the results of single molecule experiments. The outcome of these experiments are most ideally suited to reveal the description based on changes in distribution functions that characterize the conformations of proteins as the external conditions are varied.  
Other complementary theoretical viewpoints on the folding of single domain proteins have been described by several researchers \cite{OnuchicCOSB04,Shakhnovich06ChemRev,Dill08ARB,Brooks01ARPC,Pande05ARBBS}. 
\end{extract}

\section{UNIVERSALITY IN PROTEIN FOLDING THERMODYNAMICS}
\begin{extract}
 The natural variables that should control the generic behavior of protein folding are the length ($N$) of the protein, topology of the native structure \cite{BakerNature00}, symmetry of the native state \cite{Klimov05JMB,Wolynes96PNAS}, and the characteristic temperatures that give rise to the distinct ``phases" that a protein adopts as the external conditions (such as temperature $T$ or denaturant concentration ([C])) are altered \cite{Thirum95JPI}.  In terms of these variables, several universal features of the folding process can be derived, which.   shows that certain aspects of  protein folding can be understood using concepts developed in polymer physics \cite{deGennesbook,Florybook2,Edwardsbook, GrosbergBook}. 
\end{extract}
\subsection{Protein Size Depends on Length}
\begin{extract}
Under strongly denaturing conditions proteins ought to exhibit random coil characteristics. If this were the case, then based on Flory theory \cite{Florybook2}, the radius of gyration ($R_G$)  of proteins in the unfolded state must scale as  $R_G^D \approx a_DN^{\nu}$ where $a_D$ is a characteristic Kuhn length, $N$ is the number of amino acid residues,  and $\nu \approx 0.6$. Analysis of experimental data indeed confirms the Flory prediction (Fig.1a) \cite{Kohn04PNAS}, which holds good for homopolymers in good solvents.  Because folded proteins are maximally compact the native states should obey $R_G^N \approx a_NN^{\nu}$ with $\nu=1/3$.  Explicit calculations of $R_G$ for a large number of proteins in the Protein Data Bank (PDB) show that the expected scaling is obeyed for the folded states as well (Fig.1b) \cite{Dima04JPCB}.    
\end{extract}
\subsection{Characteristic Phases}
\begin{extract}
Proteins are finite-sized systems
that undergo phase changes as the quality of solvent is decreased.
As the
$T$ ([C]) is lowered to the collapse temperature $T_{\theta}$ ([C]$_{\theta}$), which decreases the solvent quality, a transition from an expanded to an ensemble of compact structures must take place.   The collapse 
transition can be either first or second order \cite{deGennesbook},
depending on the nature of the solvent-mediated interactions.  In a protein there are additional
energy scales 
that render a few of the exponentially large number of conformations
lower in free energy than the rest. These minimum energy compact structures ({\bf MECS}) direct the folding process \cite{Camacho93PRL}. 
When the temperature is lowered to the folding transition
temperature $T_F$, a transition to the folded
native structure takes place. 
These general arguments suggest
that there are minimally three phases for a protein as $T$ or [C]
is varied. They are the unfolded ($U$) states, an ensemble of intermediate
($I$) structurally heterogeneous compact states, and the native state. 

An  order parameter that distinguishes the $U$ and $I$ states is the monomer density,
$\rho  = N/R_G^3$. It follows from the differences in the size dependence of $R_G$ in the $U$ and $I$ states with $N$ (Fig. 1) that $\rho \approx 0$ in the $U$ phase
while $\rho \approx O(1)$ in the $I$ and the {\bf NBA}.
The structural overlap function ($\chi$), which measures the similarity to the native structure, is necessary to differentiate between the $I$ and the conformations in the {\bf NBA}. The collapse temperature may be estimated from the changes in the $R_G$ values of the unfolded state as $T$ is lowered while $T_F$ may be calculated from $\Delta \chi = \langle \chi^2 \rangle -\langle \chi \rangle ^2$, the fluctuations in $\chi$. 
\end{extract}
\subsection{Scaling of Folding Cooperativity with $N$ is Universal}
\begin{extract}
A hallmark of the folding transition of small single domain proteins is that it is remarkably cooperative (Fig.2). 
The marginal stability criterion can be used to infer the $N$-dependent growth of a dimensionless measure of  cooperativity $\Omega_c = \frac{T_F^2}{\Delta T}|\frac{df_{NBA}}{dT}|_{T=T_F}$ \cite{ThirumFoldDes98},  where $\Delta T$ is the full width at half maximum of $|\frac{df_{NBA}}{dT}|$,  in a way that reflects both the finite-size of proteins and the global characteristics of the denatured states.    

The dependence of  $\Omega_c$ on  $N$ is derived  using the following
arguments \cite{Li04PRL}. (i)   $\Delta \chi$  is
analogous to susceptibility in magnetic systems and hence can be written as $\Delta \chi = T|d\langle \chi \rangle/dh|$, where $h$ is an ordering field
conjugate to  $\chi$. Because  $\Delta \chi$  is dimensionless, we expect
that the ordering field $h \sim  T$. Thus, $T|d\langle \chi \rangle/dT|\sim T|df_{NBA}/dT|$ plays the role of susceptibility in magnetic systems.  (ii) Efficient folding in
apparent two-state folders implies $T_F \approx  T_{\theta}$ \cite{Camacho93PNAS} (or equivalently $C_{\theta}\approx C_F$ \cite{ThirumFoldDes98} when folding is triggered by denaturants).  Therefore, the critical exponents that
control the behavior of the polypeptide chain at $T_{\theta}$  must control the 
thermodynamics of the folding phase transition. At $T \approx  T_{\theta} \approx  T_F$ the Flory radius $R_G \sim \Delta T^{-\nu} \sim N^{\nu}$. Thus,   $\Delta T \sim N^{-1}$ (Fig.2b).  Because of the analogy to
magnetic susceptibility, we expect $T|d \langle \chi \rangle/dT| \sim  N^{\gamma}$. 
Using these results we obtain 
$\Omega_c \approx N^{\zeta}$ where $\zeta = 1 + \gamma$, which follows from the hypothesis that
$T_F \approx  T_{\theta}$. 
The fifth order $\epsilon$ expansion for polymers using $n$-component field theory with $n\rightarrow 0$ gives $\gamma =  1.22$, giving $\zeta = 2.22$ \cite{KleinertBook}. 

The linear fit to the log-log plot of the dependence of  $\Omega_c$ for proteins shows that 
$\zeta = 2.17 \pm   0.09$ for proteins (Fig.2c). 
The
remarkable finding that expresses cooperativity in terms of $N$ and $\zeta$ gives further credence to
the proposal that efficient folding is achieved if sequences are poised to have $T_F \approx T_{\theta}$ \cite{Camacho93PNAS,Klimov96Proteins}. 
\end{extract}

\section{GENERAL PRINCIPLES THAT GOVERN FOLDING KINETICS}
\begin{extract}
A few general conclusions about how proteins access the {\bf NBA} may be drawn by visualizing the folding process in terms of navigation of a large dimensional folding landscape (Fig. 3a). 
Dynamics of random heteropolymers have shown that their energy landscapes are far too rugged to be explored \cite{Bryngelson89JPC} on typical folding times (on the order of milliseconds).  Therefore,  the energy landscape 
of many evolved proteins must be  smooth (or funnel-like \cite{LeopoldPNAS92,OnuchicCOSB04,Dill08ARB}) i.e., the gradient of the energy landscape 
  towards the {\bf NBA} is ``large" enough that the biomolecule does not pause in  Competing Basins of
 Attraction ({\bf CBA}s) for long times during the folding process.  Because of energetic and topological frustration  the folding landscapes of even highly evolved proteins is rugged on length scales smaller than $R_G$ \cite{ThirumalaiACR96,Hyeon07JP}.  In the folded state, the hydrophobic residues are usually sequestered in the interior 
while polar and charged residues are better accommodated on the protein surface.
Often these conflicting requirements cannot be simultaneously satisfied and hence proteins can be energetically ``frustrated'' \cite{Guo95BP,Clementi00JMB}. 
If the packing of locally formed structures is in conflict with the global
 fold then the polypeptide chain is topologically frustrated. Thus, the energy landscape rugged 
on length scales that are larger than those in which secondary structures ($\approx (1-2)$ $nm$) form even if folding can be globally described using the two-state approximation.

There are several implications of the funnel-like and rugged landscapes for folding kinetics (Fig.3a). (i) Folding pathways are diverse. The precise folding trajectory that a given molecule follows depends on the initial conformation and the location in the landscape from which folding commences. 
(ii) If the scale of ruggedness is small compared to $k_B T$ ($k_B$ is the Boltzmann constant) then trapping in {\bf CBA}s for long times is unlikely, and hence folding follows exponential kinetics. (iii) On the other hand if the space of {\bf CBA}s is large then a substantial fraction of molecules can be kinetically trapped in one or more of the {\bf CBA}s.  If the time scale of interconversion between the conformations in the {\bf CBA}s and the {\bf NBA} is long then the global folding would occur through  well-populated intermediates. 
\end{extract}

\subsection{Multiple Folding Nuclei (MFN) Model}
\begin{extract}
Theoretical studies
\cite{Guo95BP,Thirumalai95BP,Bryngelson90BP,Shakhnovich94BC}
and some experiments \cite{Fersht95PNAS,Itzhaki95JMB} 
suggest that efficient folding of these proteins is consistent
with a NC mechanism according to which the rate limiting step involves the formation of 
 one of the folding nuclei.  Because the formation of the folding
nucleus and the collapse of the chain are nearly synchronous, we referred to this 
process as the NC mechanism. 

Simple theories  have been proposed to estimate the free energy cost of producing a structure that contains a critical number $N_R^*$ residues whose formation drives the structure to the native state \cite{WolynesPNAS97,Guo95BP,Chen08JPCB}. 
In the simple NC picture the barrier to folding occurs because the formation of contacts (native or non-native) involving the $N_R^*$ residues, while enthalpically favorable, is opposed by surface tension.  
In addition, formation of non-native interactions in the transition state also creates strain in the structures representing the critical nuclei. 
Using a version of the 
nucleation theory and structure-based thermodynamic data, we showed the average size of the most probable nucleus $N_R^*$, for single
domain proteins, is between 15-30 residues \cite{Chen08JPCB}. 

Simulations using lattice and off-lattice  established the validity of the MFN model according to which certain contacts (mostly native) in the conformations in Transition State Ensemble ({\bf TSE}) form with substantial probability ($> 0.5$).  An illustration (Fig.3c) is given from a study of the lattice model with side chains \cite{Klimov01Proteins} in which the distribution of native contacts ($P_N(q)$) shows that about 45\% of the total number of native contacts have high probability of forming in the {\bf TSE} and none of them form with unit probability.  Although important \cite{Li00NSB}, very few non-native contacts have high probability of forming at the transition state.  
\end{extract}

\subsection{Kinetic Partitioning Mechanism (KPM)}
\begin{extract}
When the scale of roughness far exceeds $k_BT$ so that the folding landscape partitions into a number of distinct {\bf CBA}s that are separated from each other and the {\bf NBA} by discernible free energy barriers (Fig.3a) then folding is best described by the KPM.  
A fraction $\Phi$  of molecules can reach
the {\bf NBA} rapidly Fig.3a).  The remaining
fraction, $1 - \Phi$, is trapped in a manifold of discrete
intermediates. 
Since the transitions from the {\bf CBA}s to
the {\bf NBA} involve partial unfolding, crossing of the free energy barriers for this class of
molecules is slow. 
The KPM explains the not only the folding of complex structured proteins but also counterion-induced assembly of RNA especially {\it{Tetrahymena}} ribozyme \cite{Hyeon05BC}. 
For RNA and large proteins $\Phi \approx (0.05 - 0.2)$ \cite{Hyeon05BC,Kiefhaber95PNAS,Li08PNAS}.   The KPM is also the basis of the Iterative Annealing Mechanism \cite{LorimerPNAS96,Tehver08JMB}.  

\end{extract}

\subsection{Three Stage Multipathway Kinetics and the Role of $N$}
\begin{extract}
The time scales associated with distinct routes followed by the unfolded molecules (Fig.3) can be approximately estimated using $N$. For the case when $\Phi \approx 0$, the folding time $\tau_F \sim\tau_0 N^{2 + \theta}$ where the $1.8 \le \theta \le 2.2$ \cite{Thirum95JPI}.  The theoretically predicted power law dependence  was validated in lattice model simulations in a subsequent study \cite{Shakhnovich96PRL}. 

Simulations using lattice and off-lattice models showed that the molecules that follow the slow track reach the native state in three stages (Fig.3b) \cite{Camacho93PNAS,Guo95BP,Thirum95JPI}.  

{\it{Non-specific Collapse:}}  In the first stage the polypeptide chain  collapses to an ensemble of  compact conformations driven  by the hydrophobic forces.   The conformations even at this stage
might have fluctuating secondary and tertiary  structures. By adopting the kinetics of coil-globule formation in hompolymers it was shown that the time scale for non-specific collapse $\tau_{nc} \approx \tau_{c0} N^2$. 

{\it{ Kinetic Ordering:}} 
In the second phase the polypeptide chain effectively discriminates between
the exponentially large number of compact conformations to attain a large fraction of
native-like contacts.  At the end of this
stage the molecule finds one of the basins corresponding to the {\bf MECS}. 
 Using an analogy to reptation in polymers we suggested that the  time associated with this stage is $\tau_{KO} \sim \tau_{KO0} N^3$ \cite{Camacho93PRL}.

{\it{All or None:}}  The final stage of folding corresponds to activated transitions from one of
the {\bf MECS} to the native state. 
A detailed analysis of several independent trajectories for both lattice and
off-lattice simulations suggest that there are multiple pathways that lead to the structures
found at the end of the second stage. There are relatively few paths that connect
the native state and the numerous native-like conformations located at the end of the
second stage (Fig.3b).  

In majority of ensemble experiments only the third folding stage is measured. The folding time $\tau_F \approx \tau_0 \exp(\Delta F^{\ddagger}/k_BT)$  where  the barrier height $\Delta F^{\ddagger} \approx \sqrt{N}$. Others have argued that $\Delta F^{\ddagger}\approx N^{2/3}$ \cite{WolynesPNAS97,FinkelsteinFD97}.
The limited range of $N$ for which data are available makes it difficult to determine the exponent unambiguously. 
However, correlation of the stability of the folded states \cite{Thirum95JPI} expressed as $Z$-score ($\propto\sqrt{N}$) with folding time \cite{Klimov98JCP} shows that $\sqrt{N}$ scaling 
\cite{Naganathan05JACS,Fernandez07BJ} is generic (Fig. 3d). 
\end{extract}

\section{MOVING FORWARD : NEW DEVELOPMENTS}
\begin{extract}
Theoretical framework and
 simulations (especially using a variety of
coarse-grained models \cite{Hill09IJMS,Clementi00JMB,GosaviJMB06,Gosavi08PNAS,OnuchicPNAS2000,Karani04PNAS,Karani03JMB,Honeycutt90PNAS}) have been instrumental in making testable predictions for folding of a number of proteins. 
For example, by combining structural analyses of a number of SH3 domains using polymer theory, and off-latice simulations we showed that the stiffness of the distal loop is the reason for the observation of polarized transition state in src SH3 and $\alpha$-spectrin SH3 \cite{Klimov02JMB}. 
The theoretical prediction was subsequently validated by Serrano and coworkers \cite{Serrano03PS}. This and other successful applications that combine simulations and experiments legitimately show that, from a broad perspective, how proteins
fold is no longer as daunting a problem as it once seemed.  

On the experimental front impressive advances, especially using single
molecule FRET (smFRET)  \cite{Nienhaus06Macro,Sherman06PNAS,Rhoades04JACS,Haran03JPCM,Rhoades03PNAS,Buscaglia05JMB,Kubelka04COSB,Lipman03Sci,
Schuler02Nature} and Single Molecule Force spectroscopy (SMFS) \cite{Block08COCB,Fernandez09PNAS,Fernandez04Science,Dietz06PNAS}  pose new challenges that 
demand more quantitative predictions. 
Although still  in their infancy, single molecule experiments have established the need to
describe folding in terms of shifts in the distribution functions of the properties of the proteins as the conditions are changed, rather than using the more traditional well-defined pathway approach.  
New models that
not only make precise connections to experiments but also produce far
reaching predictions are needed in order to take the next leap in
the theory of protein folding.
\end{extract}

\section{MOLECULAR TRANSFER MODEL (MTM) : CONNECTING THEORY AND EXPERIMENT}
\begin{extract}
Almost all of the computational studies to date have been done using
temperature to trigger folding and unfolding, while protein stability and kinetics in a majority of
the experiments are probed using chemical denaturants.  
A substantial conceptual advance to
narrow the gap  between experiments and computations was made with the
introduction of the MTM theory \cite{ObrienBIOCHEM2009,Obrien08PNAS}.  
 The goal of the MTM is to combine
simulations at
condition {\bf A},  and reweighting the
protein conformational ensemble appropriately such that the behavior of the protein under
solution condition {\bf B}($\equiv\{T_B,pH_B,[C_B]\}$) can be accurately
predicted without running additional simulations at
{\bf B}. By using  the partition function
$Z(A) = \sum_i e^{-\beta_AE_i(A)}$ in  condition \textbf{A} ($\beta_A
=(k_BT_A)^{-1}$ and $E_i(A)$ is the potential energy of the
$i^{th}$ microstate),  and  the free
energy cost of transferring $i$ from \textbf{A} to \textbf{B} (denoted
$G_{tr,i}(A\rightarrow B)$)
the partition function $Z(B) = \sum_i
e^{-\beta_B(E_i(A)+G_{tr,i}(A\rightarrow B))}$
in condition \textbf{B} can be calculated (Fig.4a).
\end{extract}

\subsection{Applications to Protein L and Cold Shock Protein}
\begin{extract}
In the applications of MTM theory to date we have used the $C_{\alpha}$-side chain model
($C_{\alpha}-SCM$) for proteins so that accurate calculation of $Z(A)$
can be made.  The phenomenological
Transfer Model \cite{Bolen08ARB}, which  accurately predicts $m$-values for
a large number of proteins (Fig.4b), is used to compute $G_{tr,i}(A\rightarrow
B)$ for each protein conformation using the measured [C]-dependent
transfer free energies of amino side chains and backbone from water to
a [C]-molar solution of denaturant or osmolyte.

The success of the MTM is evident by comparing the results of simulations with the GdmCl-dependent changes in $f_{NBA}$ and FRET efficienty ($\langle E\rangle$) for protein L and CspTm cold shock proteins (Fig.4c and 4d). 
Notwithstanding the discrepancies among different experiments, the predictions of $\langle E\rangle$ as a function of GdmCl concentration are in excellent agreement with experiments (Fig.4d).    
The calculations in Fig.4 are the first to show that quantitative agreement between theory and experiment can be obtained, thus setting the stage for extracting [C]-dependent structural changes that occur during the folding process. 
\end{extract}

\subsection{Characterization of the Denatured State Ensemble}
\begin{extract}
How does the {\bf DSE} change as [C] decreases? A total picture of the folding process requires knowledge of the distributions of various properties of interest, namely, secondary and tertiary structure contents and the end-to-end distance $R_{ee}$ as [C] changes. 
The MTM simulations reveal a number of surprising results regarding the {\bf DSE} properties of globular proteins in general and protein L and CspTm in particular. 
(i) Certain properties ($R_G$ for example) may indicate that high denaturant concentration is a good solvent for proteins (Fig.1a) while others  give a more nuanced picture of the {\bf DSE} properties \cite{Obrien08PNAS}. 
If high [C] is a good solvent then from polymer theory it can be shown that the end-to-end distribution function $P_T(x)\sim x^{\delta}\exp{(-x^{\frac{1}{1-\nu}})}$, where $x=R_{ee}/\langle R_{ee}\rangle$ ($\langle R_{ee}\rangle$ is the average end-to-end distance) should be universal with the exponent $\delta \approx 0.3$ in three dimensions.  
Although the scaling of $R_G^D\sim N^{\nu}$ of the {\bf DSE} with $\nu\approx 0.6$ (Fig.1a) suggests that the {\bf DSE} can be pictured as a  random coil, the simulated $P(x)$ for protein L deviates from $P_T(x)$, which shows that even at high GdmCl remnants of structure must persist (Fig.5a). 
(ii) An  important finding in smFRET experiments is that the statistical characteristics of the {\bf DSE} changes substantially as [C]$<C_m$, the midpoint concentration at which the populations of the unfolded and folded structure are equal. 
For a number of proteins, including protein L and CspTm, there is a collapse transition predicted theoretically (Fig.5b) and demonstrated in smFRET \cite{Schuler08COSB,Sherman06PNAS}. 
For $[C]>>C_m$ only moderate changes in $R_G^D$  are observed while larger changes occur as $[C]<C_m$ (Fig.5b). 
Concomitant with the equilibrium collapse, the fraction of residual structure increases, with the largest increase occurring below $C_m$ \cite{Obrien08PNAS}. 
Thus, the {\bf DSE} becomes compact and native-like as [C] decreases, which shows that the collapse process should be a generic step during the folding process (Fig.5b). 
\end{extract}

\subsection{Constancy of $m$-values and Protein Collapse}
\begin{extract}
A number of the smFRET experiments show that the {\bf DSE}
undergoes a continuous collapse as [C] decreases \cite{Ziv09JACS}, which implies that the
accessible surface area must also change with decreasing denaturant
concentration. 
These observations would suggest that the stability of the native state must be a non-linear
function of [C] even when $[C]>C_m$, which contradicts  a large
number of measurements, which show that free energy chages linearly with [C]. 
The apparent contradiction was addressed
using simulations and theory both of which
emphasize the polymer nature of proteins \cite{ObrienBIOCHEM2009,Ziv09JACS}. Explicit simulations of
protein L showed that the constancy of $m$-value (=
 $d\Delta G_{ND}/d [C]$ where $\Delta G_{ND}$ is the stability of the {\bf NBA} with respect to {\bf DSE}) arises because [C]-dependent surface area of
the backbone that
makes the largest contribution to $m$ does not change appreciably when
$[C] >C_m$.  In an alternative approach to the TM model, Ziv and
Haran \cite{Ziv09JACS} used polymer theory and experimental data on 12 proteins and
showed that  the $m$-value can be
expressed in terms of a [C]-dependent interaction energy and the
volume fraction of the protein in the expanded state (Fig. 5f). 

The continuous nature of the collapse transition has
also been unambiguously demonstrated in a series of studies by
Udgaonkar and coworkers \cite{Jha09PNAS,Udgaonkar08ARB,Lakshmikanth01NSB} who have shown that the collapse process (both thermodynamically and kinetically) is a continuous process, and the
description of folding as a two-state transition clearly obscures the
hidden complexity.  
\end{extract}

\subsection{Transition Midpoints are Residue-Dependent}
\begin{extract}
The obsession with the two-state description of the folding transition as [C] (or $T$) is changed, using only simple order parameters (see below), has led to molecular explanations of the origin of cooperativity without examination of the consequences of finite size effects. For instance, the van't Hoff criterion (coincidence of calorimetric enthalpy and the one extracted from fitting $f_{NBA(i)}$ to two states) and the superposition of denaturation curves generated using various probes such as SAXS, CD, and FRET are often used to assert that protein folding can be described using only two states. 
However, these descriptions, which use only a limited set of order parameters, are not adequate for fully describing the folding transition.

The order parameter theory for first and second order phase transitions is most useful when the decrease in symmetry from a disordered to an ordered phase can be described using simple physically transparent variables. For example, magnetization and Fourier components of the density are appropriate order parameters for spin systems (second order transition) and the liquid to solid transition (first order transition) \cite{Ramakrishnan79PRB}, respectively. In contrast,  
devising order parameters for complex phase transitions (spin glass transition \cite{Mezardbook} or liquid to glass transiton \cite{ThirumalaiPRA89}) is often difficult. 
A problem in using only simple order parameters in describing the folding phase transition is that the decrease in symmetry in going from the unfolded to the folded state cannot be unambiguously identified (see however \cite{Wolynes96PNAS,Klimov05JMB}). 
It is likely that multiple order parameters are required to characterize protein structures, which makes it difficult to assess the two state nature of folding using only a limited set of observables. 
Besides enthalpy and $R_G$ the extent of secondary and tertiary structure formation as [C] is changed can also be appropriate order parameters, for monitoring the folding process. 
Thus, multiple order parameters are needed to obtain a comprehensive view of the folding process. 

The MTM simulations can be used to monitor the changes in the conformations as [C] is changed using all of the order parameters described above. 
In particular, the simulations can be used to calculate $C_{m,i}$ the transition midpoint at which the $i^{th}$ residue is structured. 
For a strict two-state system $C_{m,i}=C_m$ the global transition midpoint for all $i$. 
However, several experiments on proteins that apparently fold in a two-state manner show that this is not the case \cite{Munoz06Nature,Holtzer97BiophysJ}. 
 Deviations of the melting temperatures of the individual residues from the global melting temperature were first demonstrated by 
Holtzer for 33 residue GCN4-LZK peptide \cite{Holtzer97BiophysJ}. 
In other words, the melting temperature is not unique but reflects the distribution in the enthalpies as the protein folds. 
These pioneering studies have been further corroborated by several recent experiments. 
Of particular note is the study of thermal unfolding of 40-residue BBL using two-dimensional NMR. 
The melting profile, using chemical shifts of 158 backbone and side-chains showed stunningly that the ordering temperatures are residue-dependent. 
The distribution of the melting temperatures peaked at $T\approx 305$ K, which correponds to the global melting temperature. 
However, the dispersion in the melting temperature is nearly 40 K!

The variations in the melting of individual residues are also seen in the MTM simulations involving denaturants. 
For protein L, the values of the denaturant (urea) unfolding of individual residues $C_{m,i}$ are broadly distributed with the global unfolding occurring at $\sim$ 6.6 M  (Fig.5e). 
The $C_{m,i}$ values for protein L depend not only on the nature of the residues as well as the context in which the residue is formed. 
For example the $C_{m,i}$ for Ala in the helical region of protein L is different from that in a $\beta$ strands, which implies that not all Alanines within the same protein are structurally  equivalent! 
Interestingly, the dispersion in melting temperature (Fig.5d) is less than in the $C_{m,i}$ values, which accords with the general notion that thermal folding is more cooperative than denaturant-induced transitions.
The variations in the melting temperatures (or $C_{m,i}$), which is due to the finite size of proteins, should decrease as $N$ becomes large.     
\end{extract}

\section{MECHANICAL FORCE TO PROBE FOLDING}
\begin{extract}
Single molecule force spectroscopy (SMFS), which directly probes the folding dynamics in terms of the time-dependent changes in the extension $x(t)$, 
has  altered our perspective of folding by explicitly showing the heterogeneity in the folding dynamics \cite{Fernandez09PNAS}. 
While bulk experiments  provide an understanding of gross properties, single molecule experiments can give a much clearer picture of the folding landscapes \cite{Hyeon07JP,Block06Science,Woodside06PNAS,Marqusee05Science}, diversity of folding and unfolding routes \cite{Mickler07PNAS,Li08PNAS}, and the timescales of relaxation \cite{Hyeon08PNAS,Kramers40Physica}.  SMFS studies using mechanical force are insightful because 
(i) mechanical force does not alter the interactions that stabilize the folded states and conformations in the {\bf CBA}s, 
(ii) the molecular extension $x$ that is conjugate to $f$ is a natural reaction coordinate, and 
(iii) they allow a direct determination of $x$ as a function of $t$ from which equilibrium free energy profiles and $f$-dependent kinetics can be inferred \cite{Hyeon08PNAS,Block06Science,Li08PNAS,TinocoARBBS04}.
Interpretation and predictions of the outcomes of SMFS results further illustrate the importance of theoretical concepts from polymer physics \cite{deGennesbook,Edwardsbook,Florybook2,GrosbergBook}, stochastic theory \cite{Kramers40Physica,Hanggi90RMP} and hydrodynamics. 

Initially SMFS experiments were performed by applying a constant load $r_f$ while more recently constant force is used to trigger folding.  While $f$ is usually applied at the endpoints of the molecule of interest, other points may be chosen \cite{Dietz06PNAS} to more fully explore the folding landscape of the molecule. 
Despite the sequence-specific architecture of the folded state, 
the FECs can be quantitatively described using standard polymer models. 
The analyses of FECs using suitable polymer models immediately provide the persistence length ($l_p$) and contour length ($L$) of the proteins \cite{Bustamante94SCI}.  Surprisingly, the FECs for a large number of proteins can be analyzed using the Wormlike Chain (WLC) for which equilibrium force as a function of extension is \cite{Marko95Macro} $l_pf/k_B T=x/L+1/4(1-x/L)^2-1/4$, with $L$ the length of the chain and $l_p$ the persistence length, the characteristic length scale of bending in the polymer.  Disruption of internal structure,  leading to rips in the FEC, provides glimpses into order of force-induced provided the structure of the folded state is known \cite{Visscher99Nature,Bustamante01Sci,Bustamante03Science}.

If $f$ is constant  using the force-clamp method \cite{Bustamante01Sci,Block03PNAS,Fernandez04Science,Visscher99Nature}, 
$x(t)$ exhibits discrete jumps among accessible basins of attractions as a function of time. 
From a long time-dependent trajectory $x(t)$ the transition rates between the populated basins can be directly calculated. 
If the time traces are ``sufficiently" long to ensure that protein ergodically samples the accessible conformations an equilibrium $f$-dependent free energy profile ($F(x)$) can be constructed \cite{Block06Science,Hyeon08PNAS}.
\end{extract}

\subsection{Transition State Location and Hammond Behavior}
\begin{extract}
If $r_f$ is a constant the force required to unfold proteins varies stochastically, which implies that the rupture force (value of $f$ at which {\bf NBA}$\rightarrow$stretched transition occurs) distribution, $P(f)$, can be constructed using a multiple measurements. 
If unfolding is described by the Bell equation (unfolding rate $k(f)=k(f=0)\exp{(f\Delta x_{TS}/k_BT)}$ 
where $\Delta x_{TS}$ is the location of the TS with respect to {\bf NBA}) then using $f^*\sim k_BT/\Delta x_{TS}\cdot\log{r_f}$, $\Delta x_{TS}$ can be estimated. 
When the response of proteins over a large range of $r_f$ is examined the $[\log{r_f},f^*]$ curve is non-linear, which is due to the dependence of $\Delta x_{TS}$ on $r_f$ \cite{Dudko03PNAS,Dudko06PRL,Dudko08PNAS,Hyeon07JP} or the presence of multiple free energy barriers \cite{EvansNature99}. 
For proteins  ($r_f\sim 100-1000$ pN/s), the value of $\Delta x_{TS}$ is in the range of $2-7$ \AA\ depending on $r_f$ \cite{RiefJMB05,Dietz04PNAS}. 

The TS movement as $f$ or $r_f$ increases, can be explained using the Hammond postulate, which states that the TS resembles the least stable species along the folding reaction \cite{HammondJACS53}. The stability of the {\bf NBA} decrease as $f$ increases, which implies that $\Delta x_{TS}$ should decrease as $f$ is increased \cite{Hyeon07JP}. 
For soft molecules such as proteins and RNA, $\Delta x_{TS}$ always decreases with increasing $r_f$ and $f$. The positive curvature in $[\log{r_f},f^*]$ plot is the signature of the classical Hammond behavior  \cite{HyeonBJ07}. 
\end{extract}

\subsection{Roughness of the Energy Landscape}
\begin{extract}
Hyeon and Thirumalai (HT) showed theoretically that if $T$ is varied in SMFS studies then the $f$-dependent unfolding rate is given by $\log k(f,T)=a+b/T-\epsilon^2/(k_BT)^2$ \cite{Hyeon03PNAS,Hyeon07JP}. 
From the temperature dependence of $k(f,T)$ (or $k(r_f,T)$) the values of $\epsilon$ for several systems have been extracted \cite{Reich05EMBOrep,RiefJMB05,Janovjak07JACS}. 
Nevo et al. measured $\epsilon$ for a protein complex consisting of nuclear receptor importin-$\beta$ (imp-$\beta$) and the Ras-like GTPase Ran that is loaded with non-hydrolysable GTP analogue. 
The values of $f^*$ at three temperatures (7, 20, 32 $^oC$) were used to obtain $\epsilon\approx [5-6] k_BT$ \cite{Reich05EMBOrep}. 
Recently, Schlierf and Rief (SR) \cite{RiefJMB05} analyzed the unfolding force distribution (with $r_f$ fixed) of a single domain of \emph{Dictyostelium discoideum filamin} (ddFLN4) at five different temperatures to infer the underlying one dimensional free energy surface. By adopting the HT theory \cite{Hyeon03PNAS} SR showed that the data can be fitted using $\epsilon=4 k_BT$ for ddFLN4 unfolding.
\end{extract}


\subsection{Unfolding Pathways from FECs}
\begin{extract}
The FECs can be used to obtain the unfolding pathways. From FEC alone it is only possible to provide a global picture of $f$-induced unfolding. 
Two illustrations, one (GFP) for which predictions preceded experiments and the other (RNase-H), illustrate the differing response to force.   
\end{extract}

\subsection{RNase-H under Tension}
\begin{extract}  
Ensemble experiments had shown that RNase-H, a 155 residue proteins, folds through an intermediate ($I$) that may be either on- or off-pathway \cite{Roder97COSB,BaldwinTIBS99II}. 
The FEC obtained from LOT experiments \cite{Marqusee05Science} showed that there is one rip in the unfolding at $f\approx 15-20$ pN corresponding to {\bf NBA}$\rightarrow U$ transition  (see Fig.6).  Upon decreasing $f$ there is a signature of $I$ in the FEC corresponding to a partial contraction in length at $f\approx 5.5$ pN, the midpoint at which $U$ and $I$ are equally populated. 
The reason for the absence of the intermediate in the unfolding FEC is due to the shape of energy landscape. 
Once the first barrier, which is significantly larger than the mechanical stability of the $I$ state relative to $U$, is crossed, global unfolding occurs in a single step.
In the refolding process, the $I$ state is reached from $U$ since the free energy barrier between $I$ and $U$ is relatively small. 
The pathways inferred from FEC is also supported by the force-clamp method. 
Even when $f$ is maintained at $f=5.5$ pN, the molecule can occasionally reach the $N$ state by jumping over the barrier between $N$ and $I$, which is accompanied by an additional  contraction in the extension.  
However, once the $N$ state is reached, RNase-H has little chance to hop back to $I$ within the observable time.    
Because in majority of cases the $I\rightarrow N$ transiton out of the {\bf NBA} ceases, it was surmised that $I$ must be on-pathway. 
\end{extract}

\subsection{Pathway Bifurcation in the Forced-Unfolding of Green Fluorescence Protein (GFP)}
\begin{extract}
The nearly 250 residue Green Fluorescence Protein (GFP) has a barrel shaped structure consisting of 11 $\beta$-strands with one $\alpha$-helix at the N-terminal.
Mechanical response of GFP, which depends both on loading rate and the stretching direction \cite{Mickler07PNAS,Dietz06PNAS}, is intricate. 
The unfolding FEC for GFP inferred from the first series of AFM experiments showed clearly well populated intermediates, which is in sharp contrast to RNAase-H. 
The assignment of the intermediates associated with the peaks in the FECs was obscured by the complex architecure of GFP. 
In the original studies \cite{Dietz06PNAS} it was suggested that unfolding occurs sequentially with the single pathway being 
$N\rightarrow[\mathrm{GFP}\Delta\alpha]\rightarrow[\mathrm{GFP}\Delta \alpha\Delta\beta]\rightarrow U$, where $\Delta \alpha$ and $\Delta\beta$ denote rupture of $\alpha$-helix and a $\beta$-strand (Fig.7) from the N-terminus \cite{Dietz04PNAS}. 
After the $\alpha$-helix is disrupted, the second rip is observed 
due to the unraveling of $\beta 1$ or $\beta 11$, both of which have identical number of residues. 
A much richer and a complex landscape was predicted using the Self-Organized Polymer model (SOP) simulations performed at the loading rate used in AFM experiments \cite{Hyeon06Structure}. 
The simulations predicted that after the formation of [GFP$\Delta\alpha$] there is a bifurcation in the unfolding pathways. 
In the majority of cases the route to the $U$ state involves population of two additional intermediates, [GFP$\Delta\beta_1$] ($\Delta \beta_1$ is the N-terminal $\beta$-strand) and [GFP$\Delta\alpha\Delta\beta_1\Delta\beta_2\Delta\beta_3$]. 
The most striking prediction of the simulations is that there is only one intermediate in unfolding pathway, $N\rightarrow[\mathrm{GFP}\Delta\alpha]\rightarrow[\mathrm{GFP}\Delta\alpha\Delta\beta_{11}]\rightarrow U$! \cite{Hyeon06Structure}. 
The  predictions along with the estimate of the magnitude of forces were quantitatively validated using SMFS experiments \cite{Mickler07PNAS}. 
\end{extract}

\subsection{Refolding upon Force-Quench}
\begin{extract}
Two novel ways of initiating refolding using mechanical force have been reported. 
In the first case a large constant force was applied to poly-ubiquitin (poly-Ub) to prepare a fully extended ensemble. 
These experiments (fig. 8a), which were the first to use $f_S\rightarrow f_Q$ jump to trigger folding, provided insights into the folding process that are in broad agreement with theoretical predictions. 
(i) The time dependent changes in $x(t)$, following a $f_S\rightarrow f_Q$ quench, occurs in at least three distinct stages. 
There is a rapid initial reduction in $x(t)$, followed by a long plateau in which $x(t)$ is roughly a constant. 
The acquisition of the native structure in the last stage, which involves two phases,  occurs in a cooperative process. 
(ii) There are large molecule-to-molecule variations in the dynamics of $x(t)$ \cite{Klimov99PNAS}.
(iii) The time scale for collapse, and folding is strongly dependent on $f_Q$ for a fixed $f_S$. 
Both $\tau_F(f_Q)$ and the $f_Q$-dependent collapse time increase as $f_Q$ increases. The value of $\tau_F(f_Q)$ can be nearly an order of magnitude greater than the value at $f_Q$.

The interpretation of the force-quench folding trajectories is found by examining (Fig.8b) the nature of the initial structural ensemble \cite{FernandezTIBS99,Li06PNAS,HyeonMorrison09PNAS}. 
The initial structural ensemble for the bulk measurement is thermally denatured ensemble ({\bf TDE}) while the initial structural ensemble under high tension is the fully stretched ({\bf FDE}, force denatured ensemble). 
Upon force quench a given molecule goes from a small entropy state ({\bf FDE}) to an ensemble with increased entropy to the low entropy folded state ({\bf NBA}) (Fig.8b). 
Therefore, it is not unusual that the folding kinetics upon force quench is vastly different from the bulk measurements. 

The folding rate upon force-quench is slow relative to bulk measurements. 
A comprehensive theory of the generic features of $x(t)$ relaxation and sequence-specific effects for folding upon force quench showed that refolding pathways and $f_Q$-dependent folding times are determined by an interplay of $\tau_F(f_Q)$ and the time scale, $\tau_Q$, in which $f_S\rightarrow f_Q$ quench is achieved \cite{HyeonMorrison09PNAS}. 
If $\tau_Q$ is small then the molecule is trapped in force-induced metastable intermediates ({\bf FIMI}s) that are separated from the {\bf NBA} by a free energy barrier. 
The formation of {\bf FIMI}s is generic to the force-quench refolding dynamics of any biopolymer.  
Interestingly, the formation of DNA toroid under tension revealed using optical tweezers experiments is extremely slow ($\sim$ 1 hour at $f_Q\approx 1$ pN). 
\end{extract}

\subsection{Force Correlation Spectroscopy (FCS)}
\begin{extract}
The relevant structures that guide folding from stretched state may be inferred using Force Correlation Spectroscopy (FCS) \cite{Barsegov05PRL}. 
In such experiments the duration $\Delta t$ in which $f_Q$ is held constant (to initiate folding) is varied (Fig.9a). 
If $\Delta t/\tau_F(f_Q)\gg 1$ then it corresponds to the situation probed in \cite{Fernandez04Science} whereas in the opposite limit folding is disrupted. 
Thus, by cycling between $f_S$ and $f_Q$, and varying the time in $f_Q$, the nature of the collapsed conformations can be unambiguously discerned. 
The theoretical suggestion was implemented in a remarkable experiment by Fernandez and coworkers using poly-Ub \cite{Fernandez09PNAS}. 
By varying $\Delta t$ from about 0.5 s to 15 s, they found that the increase in the extension upon $f_Q\rightarrow f_S$ jump could be described using a sum of two exponential functions (Fig.9b). 
The rate of the fast phase, which amounts to disruption of collapsed structures, is 40 times greater than in the slow phase that corresponds to unfolding of the native structure. 
The ensemble of mechanically weak structures that form on a $ms$ time scale corresponds to the theoretically predicted {\bf MECS}. 
The experiments also verified that {\bf MECS} are separated from {\bf NBA} by free energy barriers. 
The single molecule force clamp experiments have unambiguously showed that folding occurs by a three stage multipathway approach to the {\bf NBA}. 
Such experiments are difficult to perform by triggering folding by dilution of denaturants because $R_G$ of the {\bf DSE} is not significantly larger than the native state. 
Consequently, the formation of {\bf MECS} is far too rapid to be detected. 
The use of $f$ increases these times, making the detection of {\bf MECS} easier. 
\end{extract}

\section{CONCLUSIONS}
\begin{extract}
The statistical mechanical perspective and advances in experimental
techniques have revolutionized our view of how simple single
domains proteins fold.  What a short while ago seemed to be mere
concepts are starting to be realized experimentally thanks to the
ability to interrogate the folding routes one protein molecule at a
time.  In particular, the use of force literally allows us to place a
single protein at any point in the multidimensional free energy
surface and watch it fold.  Using advances in theory, and simulations
it appears that we have entered an era in which detailed comparisons
between predictions and experiments can be made.  Computational
methods have even been able to predict the conformations explored by
interacting proteins with the recent story of the Rop dimer being a
good example \cite{Gambin09PNAS}.
The promise that all atom simulations can be used to fold at least
small proteins, provided the force-fields are reliable,  will lead to an
unprecedented movie of the folding process that will also include the role
water plays in guiding the protein to the {\bf NBA}.

Are the successes touted here and elsewhere cause for celebration or
should it be deemed ``irrational exuberance"? It depends on what is
meant by success. There is no doubt that an edifice has been built to
rationalize and in some instances even predict the outcomes of
experiments on how  small (less than about 100 residue) proteins fold.
However, from the perspective of an expansive view of the protein
folding problem, advertised in the Introduction, much remains to be
done.  We are far from being able to predict the sequence of events
that drive the unfolded proteins to the {\bf NBA} without knowing the
structure of the folded state. From this view point both structure
prediction and folding kinetics are linked. Regardless of the level of
optimism (or pessimism) it is clear that the broad framework that has
emerged by intensely studying the protein folding
problem will prove useful as we start tackling more complex problems
of cellular functions that involve communication between a number of
biomolecules.  An example where this approach is already evident is in
the development of the Iterative Annealing Mechanism for describing  of the
function of the GroEL machine, which combines concepts from protein
folding and allosteric transitions that drive GroEL through a complex
set of conformational changes during a reaction cycle \cite{LorimerPNAS96}. Surely, the
impact of the concepts developed to understand protein folding will
continue to grow in virtually all areas of biology.
\end{extract}

\section{SUMMARY POINTS}
\begin{extract}
\begin{enumerate}
\item

Several properties of proteins
ranging from their size and  folding cooperativity  depend in an
universal manner on the number ($N$) of amino acid residues. 
The precise dependence on these properties as $N$ changes can be predicted accurately
using polymer physics concepts.

\item
Examination  of the folding landscapes leads to a
number of scenarios for self-assembly. Folding of proteins with simple
architecture can be described using the nucleation-collapse mechanism with
multiple folding nuclei, while those with complex folds reach the
Native Basin of Attraction {\bf NBA} by the Kinetic Partitioning
Mechanism.

\item
The time scales for reaching the {\bf NBA}, which occurs in three
stages depending on the protein fold, can be estimated in terms of
$N$. The predictions are well supported by experiments.

\item
The Molecular Transfer Model, which combines simulations and the
classical transfer model,  accurately predicts denaturant-dependent
quantities measured in ensemble and single molecule FRET experiments.
In the process  it is shown that the melting temperatures are
residue-dependent, which accords well with a number of experiments.

\item
The heterogeneity in the unfolding pathways, predicted theoretically,
is revealed in experiments that use mechanical force to trigger
 folding and unfolding.  Studies on GFP show the need to
combine simulations and AFM experiments to map the folding routes. Novel force protocol, proposed using theory, reveals the
presence of Minimum Energy Compact Structures predicted using simulations.

\end{enumerate}
\end{extract}

{\bf ACKNOWLEDGMENTS}

We are grateful to Valerie Barsegov, Carlos Camacho, Jie Chen, Margaret Cheung, Ruxandra Dima, Zhuyan Guo, Gilad Haran, Dmitry Klimov, Alexander Kudlay, Zhenxing Liu, George Lorimer, David Pincus, Govardhan Reddy, Riina Tehver, and Guy Ziv for collaborations on various aspects of protein folding. Support from the National Science Foundation over a number of years for our research is gratefully acknowledged.

\clearpage
{\bf Figure Legends}

{\bf Figure 1:}
(a) Dependence of $R_G^D$  on $N$.  Data are taken from \cite{Kohn04PNAS} and the line is the fit to the Flory theory.  (b) $R_G^N$ versus $N$ \cite{Dima04JPCB}. 

{\bf Figure 2:}
(a)Temperature  (in Centigrade) dependence of  $f_{NBA}$, and its derivative $|\frac{df_{NBA}}{dT}|$.  (b) Plot of $\log(\Delta T/T_F)$ versus $\log N$.  The linear fit (solid line) to the
experimental data for 32 proteins shows $\frac{\Delta T}{T_F} \sim N^{-\lambda}$ with $\lambda = 1.08 \pm 0.04$ with a correlation coefficient 0.95 \cite{Li04PRL}.
(c) Plot of $\log{\Omega_c}$  versus $\log N$.  The solid line is a fit to the data
with $\zeta =  2.17 \pm  0.09$ (correlation coefficient  is 0.95). Inset shows denaturation data.

{\bf Figure 3:}
Schematic of the rugged folding landscape of proteins with energetic and topological frustration. A fraction $\Phi$ of unfolded molecules follow the fast track (white) to the {\bf NBA} while the remaining fraction ($1-\Phi$) of slow trajectories (green) are trapped in one of the {\bf CBA}s. (b) Summary of  the mechanisms by which proteins reach their native state. The upper path is for fast track molecules.  $\Phi \approx 1$ implies the folding landscape is funnel-like.  The lower routes are  
for slow folding trajectories (green in (a)).  The number
of conformations explored in the three stages as a function of $N$ are given below, with numerical estimates for $N = 27$.  The last line 
gives the time scale for the three processes for $N =100$ using the estimates described in the text. (c) Multiple folding nuclei model for folding of a lattice model with side chains with $N = 15$ \cite{Klimov01Proteins}.  The probability of forming the native contacts (20 in the native state shown in black) in the {\bf TSE} is given in  purple. The average structure in the three major clusters in the {\bf TSE} are shown. There is a non-native contact in the most probable cluster (shown in the middle). The native state is on the right. (d) Dependence of the folding times versus $\sqrt{N}$ for 69 residues (adapted from \cite{Naganathan05JACS}). Red line is a linear fit (correlation coefficient is 0.74)  and the blue circles are data. 

{\bf Figure 4:}
(a) Diagram for MTM theory (Top).   $E_i(A)$ ($E_i(B)$) is the energy of the $i^{th}$ microstate in condition  \textbf{A} (\textbf{B}), while $Z(A)$ and $Z(B)$ are the corresponding partition functions.  (b) Linear correlation between calculated (using the TM) and measured  $m$-values for  proteins in urea \cite{Bolen07PNAS}.  (c) Predictions using MTM versus experiments (symbols). Protein L is in blue  and red is for CspTm. (d) Comparison of the predicted FRET efficiencies versus experiments for protein L. MTM results for $\langle E \rangle$  of the native state (red line), DSE (blue line),
and average (black line) are shown. Experimental values for the $\langle E \rangle$ for the {\bf DSEs} are in  blue solid squares \cite{Sherman06PNAS} and open squares \cite{Merchant07PNAS}.

{\bf Figure 5:}
(a) Distribution of $R_{ee}/\langle R_{ee} \rangle$ for the {\bf DSE} ensemble at 5, 7, and 9M GdmCl concentrations. Line is the universal curve for polymers in good solvent. (b) Predicted values of the  average $R_G$ (open circle) and $R_{ee}(x)$ as a function of urea for protein L.  The broken lines show the corresponding values for the {\bf DSE} as a function [C].   (c) Histogram of $T_{m,i}$ values for 158 protons for BBL obtained using NMR (taken from \cite{Munoz06Nature}). (d) Predicted $T_{m,i}$ values using MTM for protein L. (e) Histogram of $C_{m,i}$ values for protein L.  (f) Mean field interaction energy for three proteins versus [C] \cite{Ziv09JACS}. 

{\bf Figure 6:}
Top left shows a schematic of the LOT set up used to generate FEC and $x(t)$ for RNase-H. The curves below are unfolding FECs.  The refolding FEC on the right shows the $U \rightarrow I$ transition. The right figure shows the proposed folding landscape for the transition from $U$ to $N$ through $I$. The folding trajectory is superimposed on top of the folding landscape. Figure adapted from \cite{Marqusee05Science}.

{\bf Figure 7:}
Folding landscape for GFP obtained using SOP simulations and AFM experiments.  Top and bottom left show the folded structure and the connectvity of secondary structural elements.  The right shows the bifurcation in the pathways from the NBA to the stretched state.  

{\bf Figure 8:}
(a) Force quench refolding trajectory of poly-Ub generated by AFM (from \cite{Fernandez04Science }). The black curve shows contraction in $x(t)$ after fully stretching poly-Ub. (b) Sketch of the folding mechanism of a polypeptide chain upon $f_S \rightarrow f_Q$ quench. Rapid quench generates a plateau in $x(t)$ ({\bf FIMI}) followed by exploration of {\bf MECS} prior to reaching the {\bf NBA}. Chain entropy goes from a small value (stretched state) to large value (compact conformations) to a low value ({\bf NBA}).

{\bf Figure 9:}
(a) Sketch of force pulse used in Force Correlation Spectroscopy (FCS). Polypeptide chain is maintained at $f_Q$ for arbitrary times before stretching. 
(b) Increase in extension of poly-Ub upon application of stretching force for various $\Delta t$ values \cite{Fernandez09PNAS}.

\clearpage

\begin{figure}
  \centering
  \begin{center}$
  \begin{array}{cc}
  \subfloat[]{\label{}\includegraphics[width=0.48\textwidth]{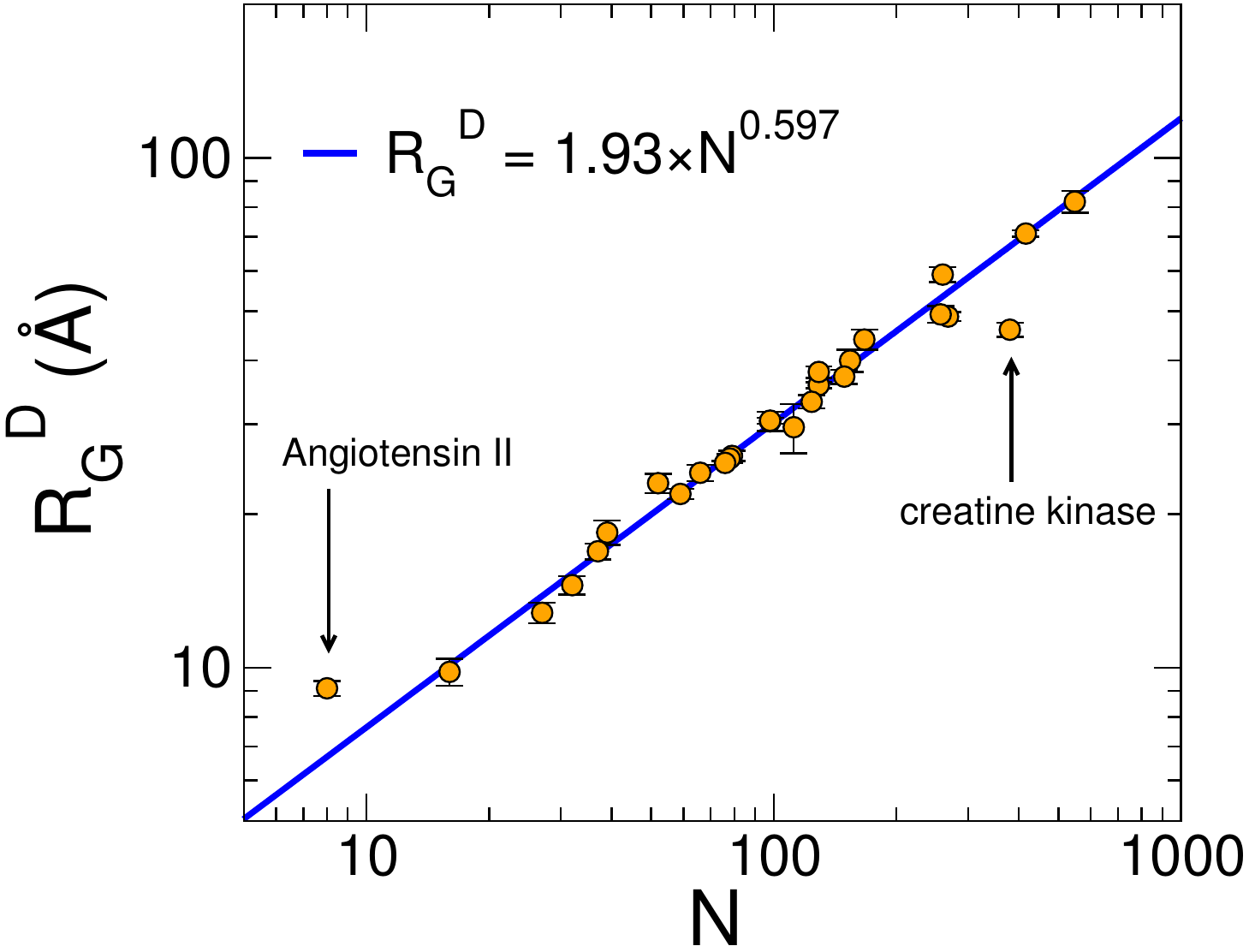}}&       
  \qquad         
  \subfloat[]{\label{}\includegraphics[width=0.46\textwidth]{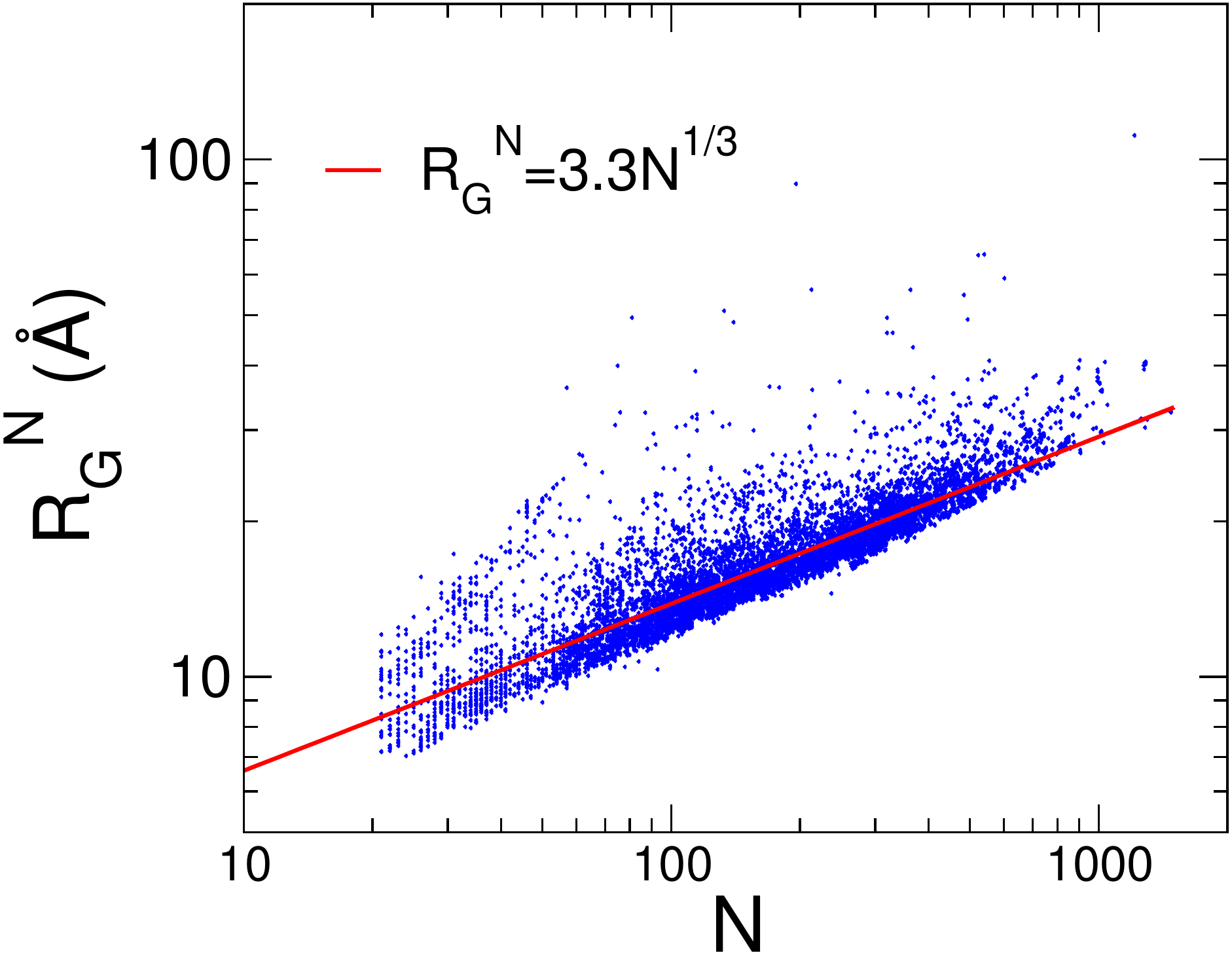}}
  \end{array}$
  \end{center}
 \caption{}
  \label{}
\end{figure}
\clearpage
\begin{figure}
\begin{center}
  \subfloat[]{\label{}\includegraphics[width=0.5\textwidth]{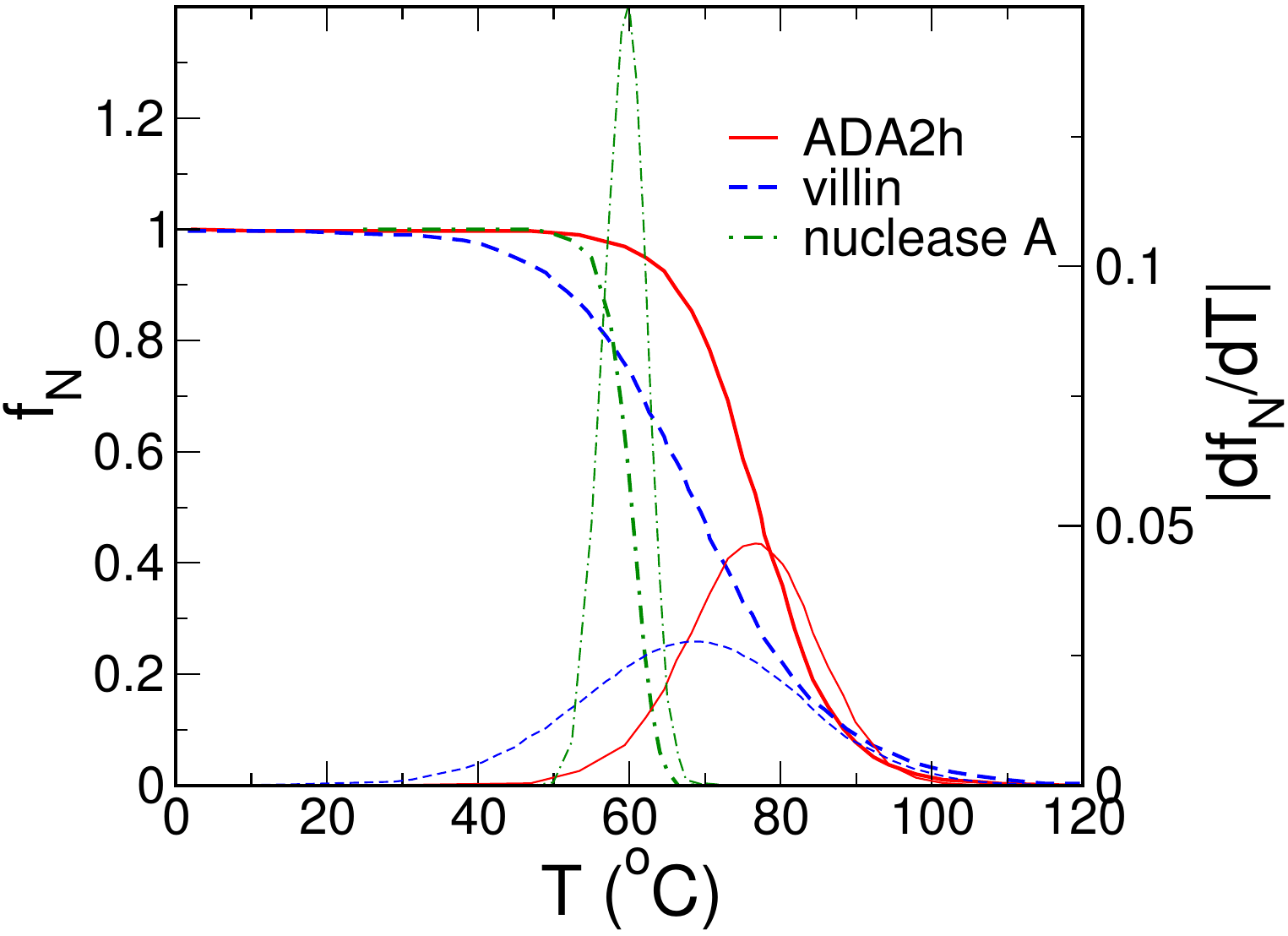}}
 \end{center}
  \begin{center}$
  \begin{array}{cc}
  \subfloat[]{\label{}\includegraphics[width=0.4\textwidth]{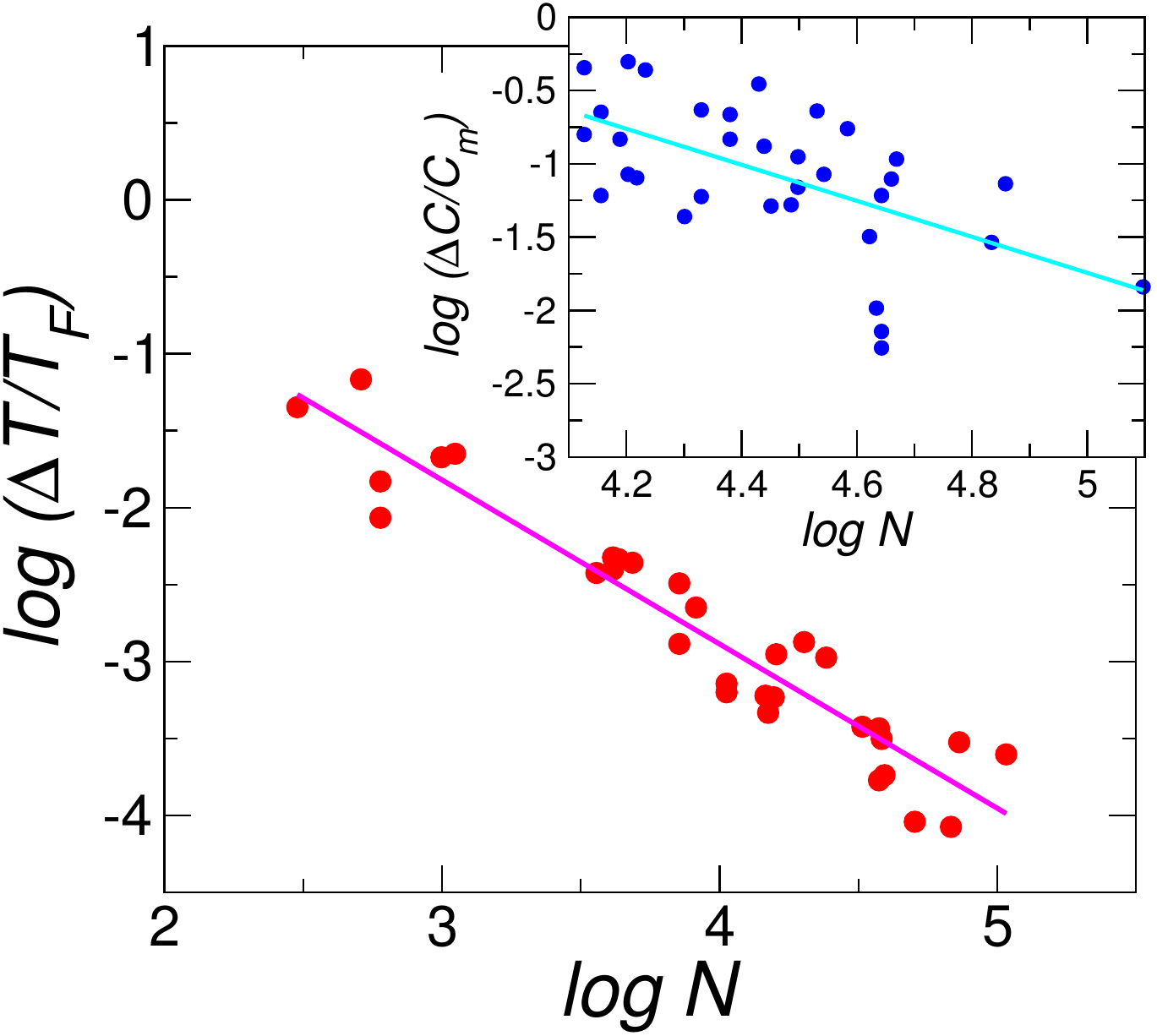}}&
  \subfloat[]{\label{}\includegraphics[width=0.4\textwidth]{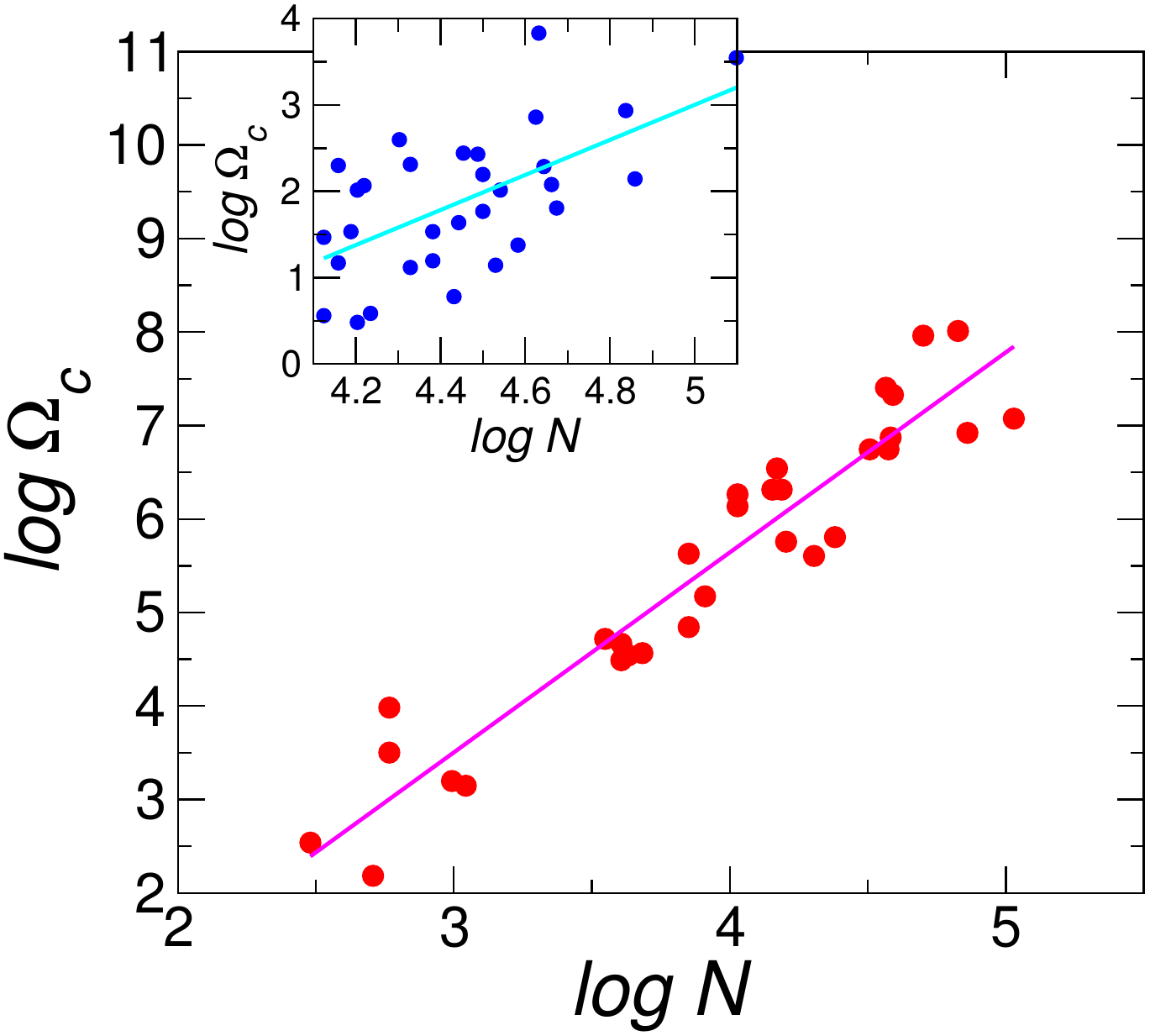}}
  \end{array}$
  \end{center}
 \caption{}
  \label{}
\end{figure}

\begin{figure}
  \begin{center}$
  \begin{array}{cc}
  \subfloat[]{\label{}\includegraphics[width=0.5\textwidth]{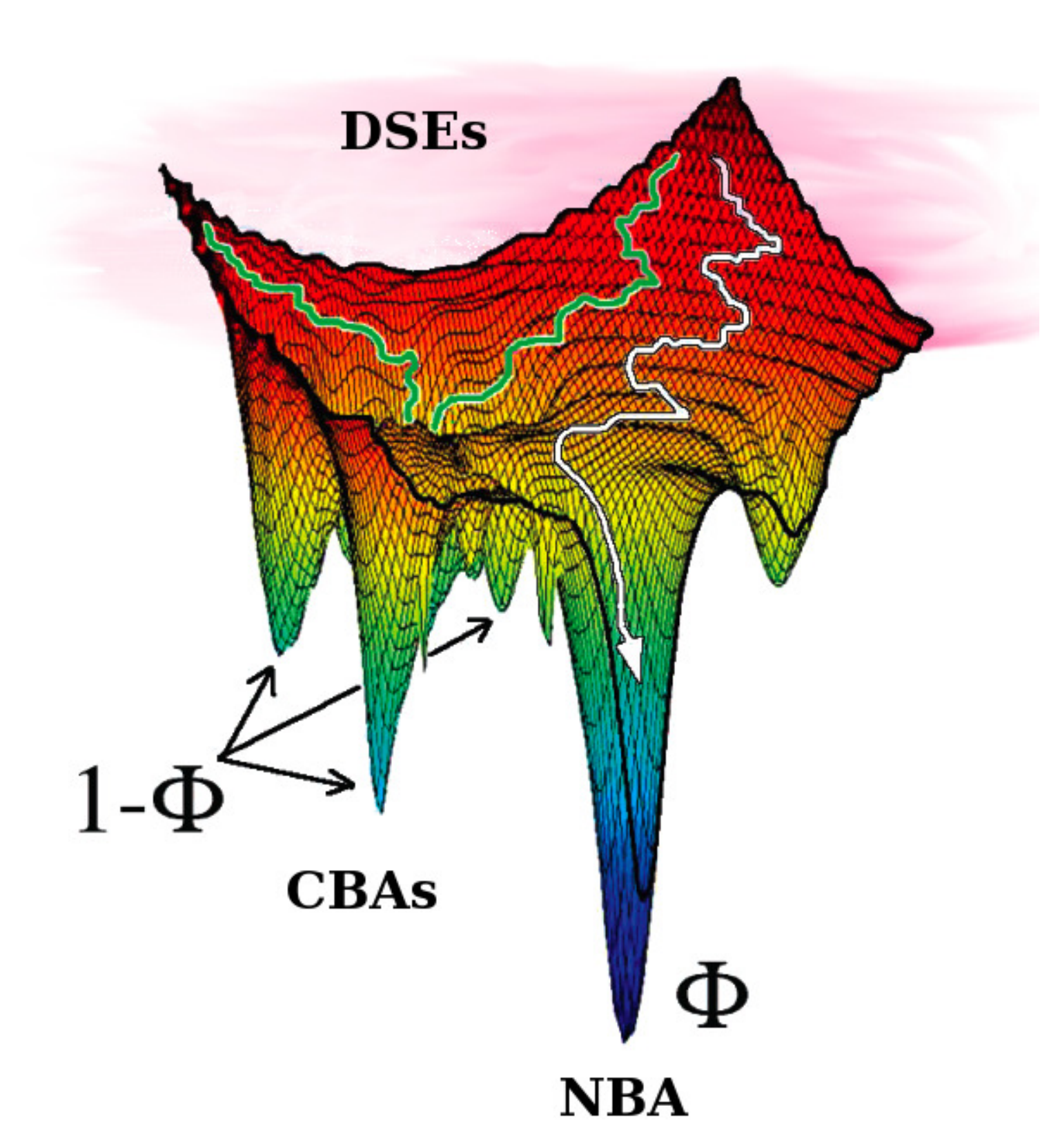}}&    
  \subfloat[]{\label{}\includegraphics[width=0.72\textwidth]{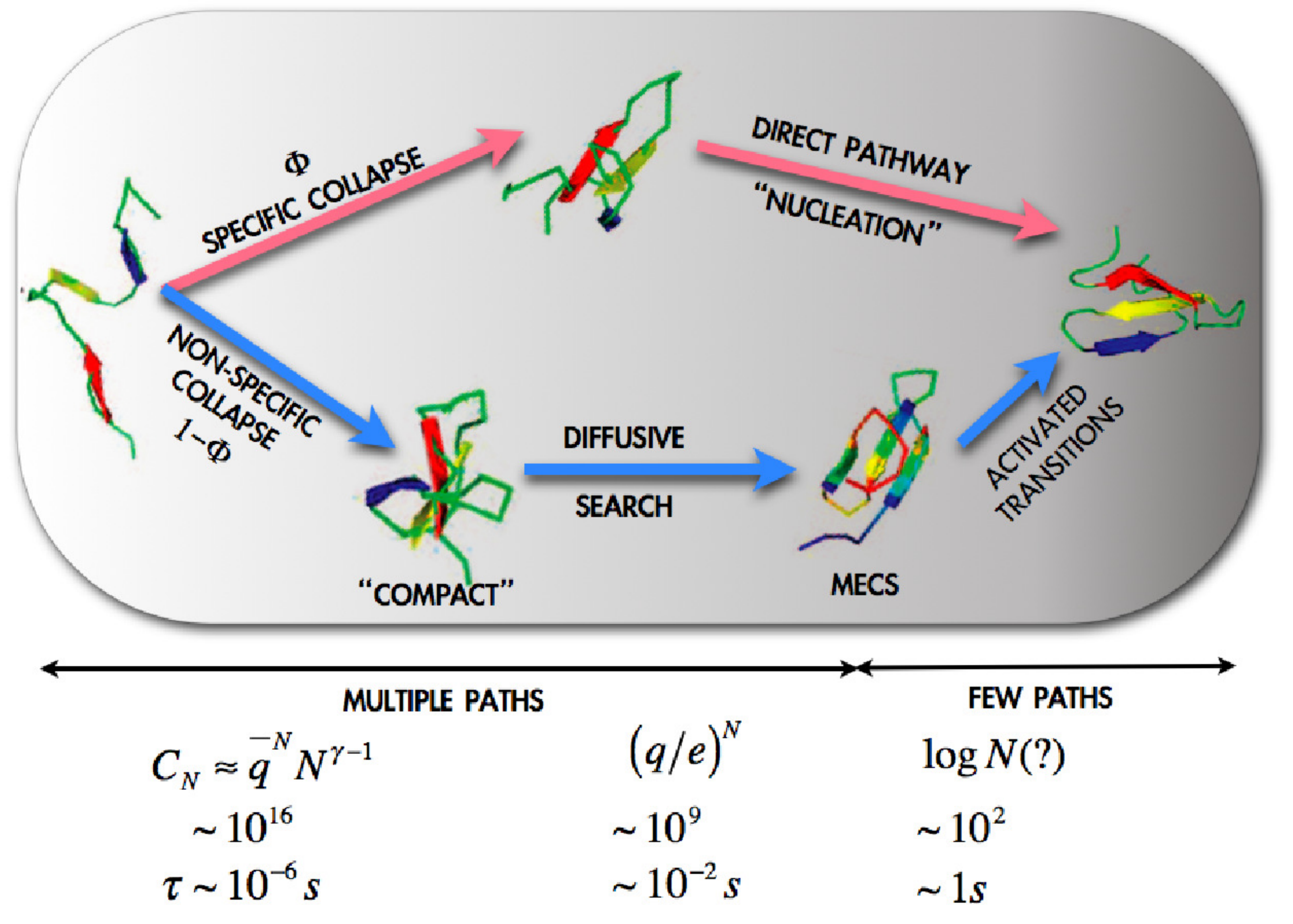}}\\
  \subfloat[]{\label{}\includegraphics[width=0.6\textwidth]{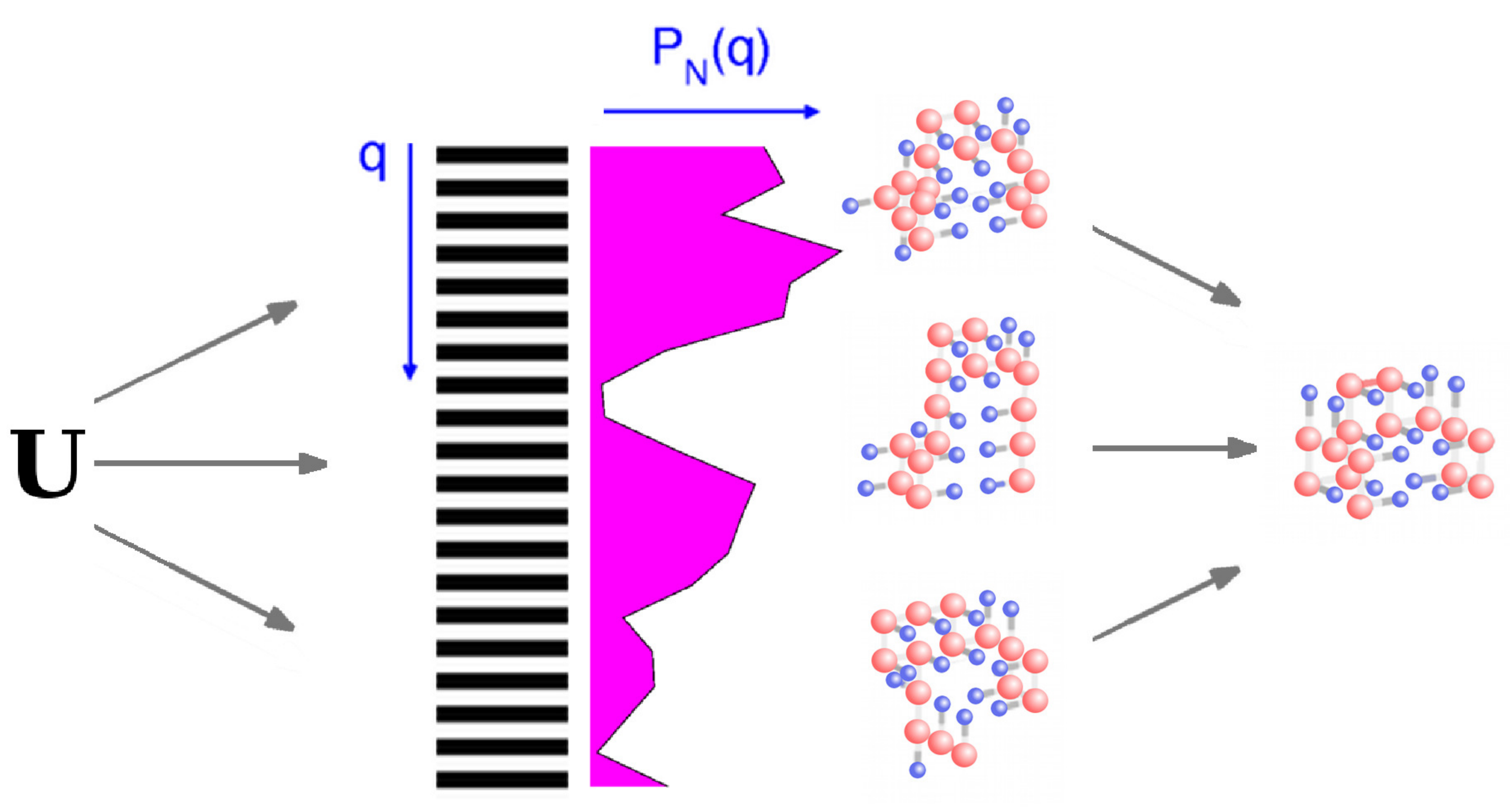}}&
  \subfloat[]{\label{}\includegraphics[width=0.4\textwidth]{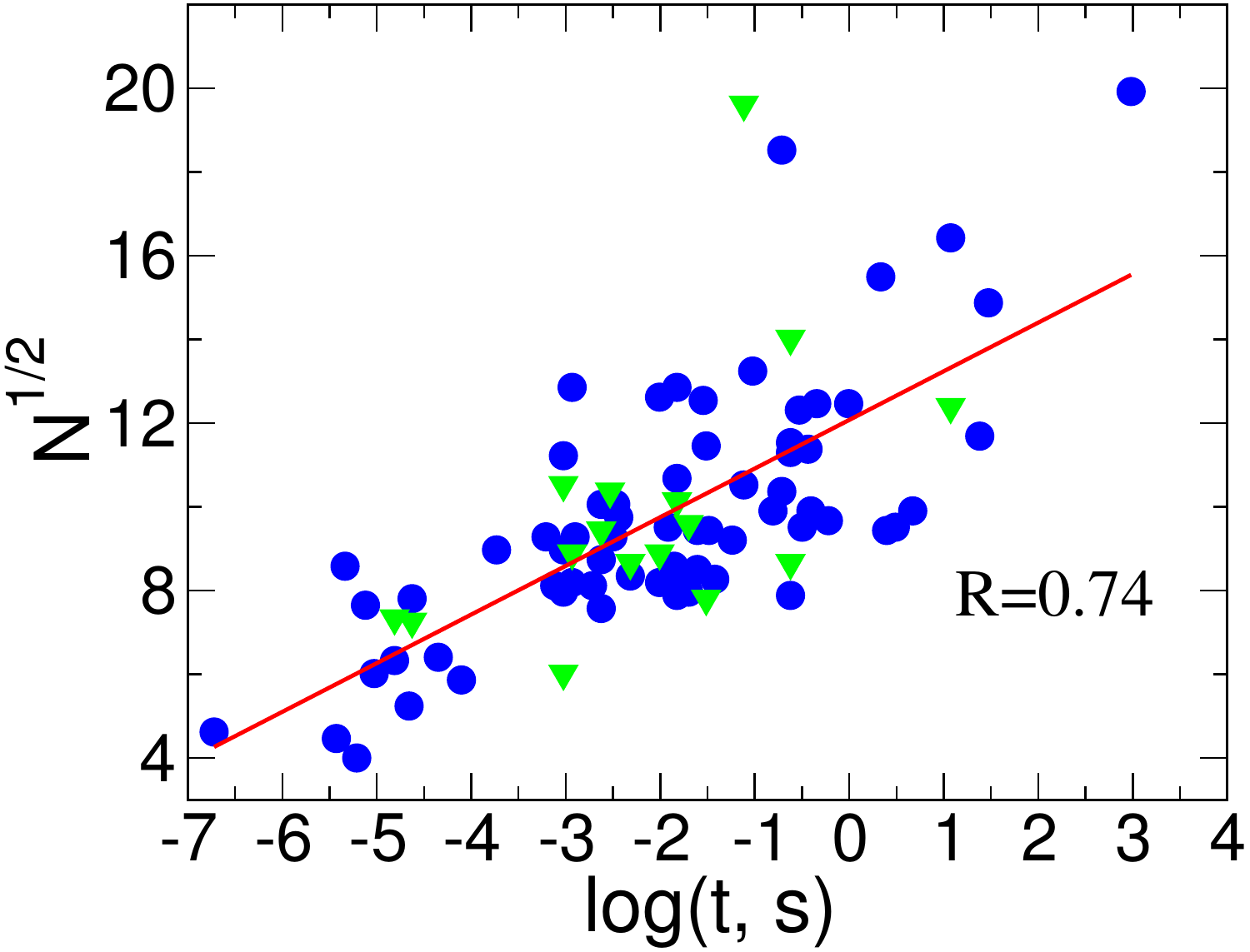}}
  \end{array}$
  \end{center}
 \caption{}
  \label{}
\end{figure}

\begin{figure}
  \begin{center}$
  \begin{array}{cc}
  \subfloat[]{\label{}\includegraphics[width=0.4\textwidth]{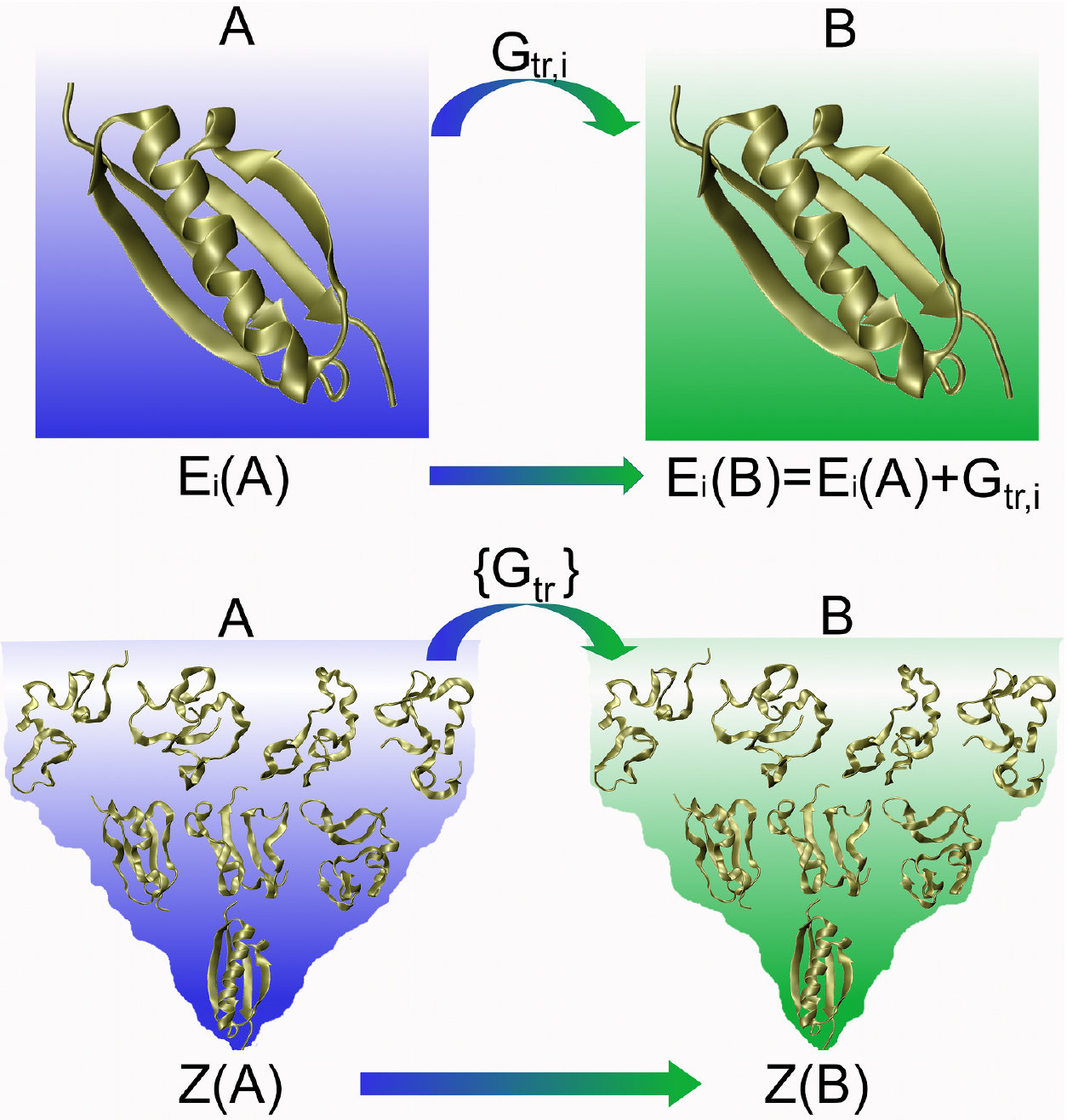}}&    
  \subfloat[]{\label{}\includegraphics[width=0.4\textwidth]{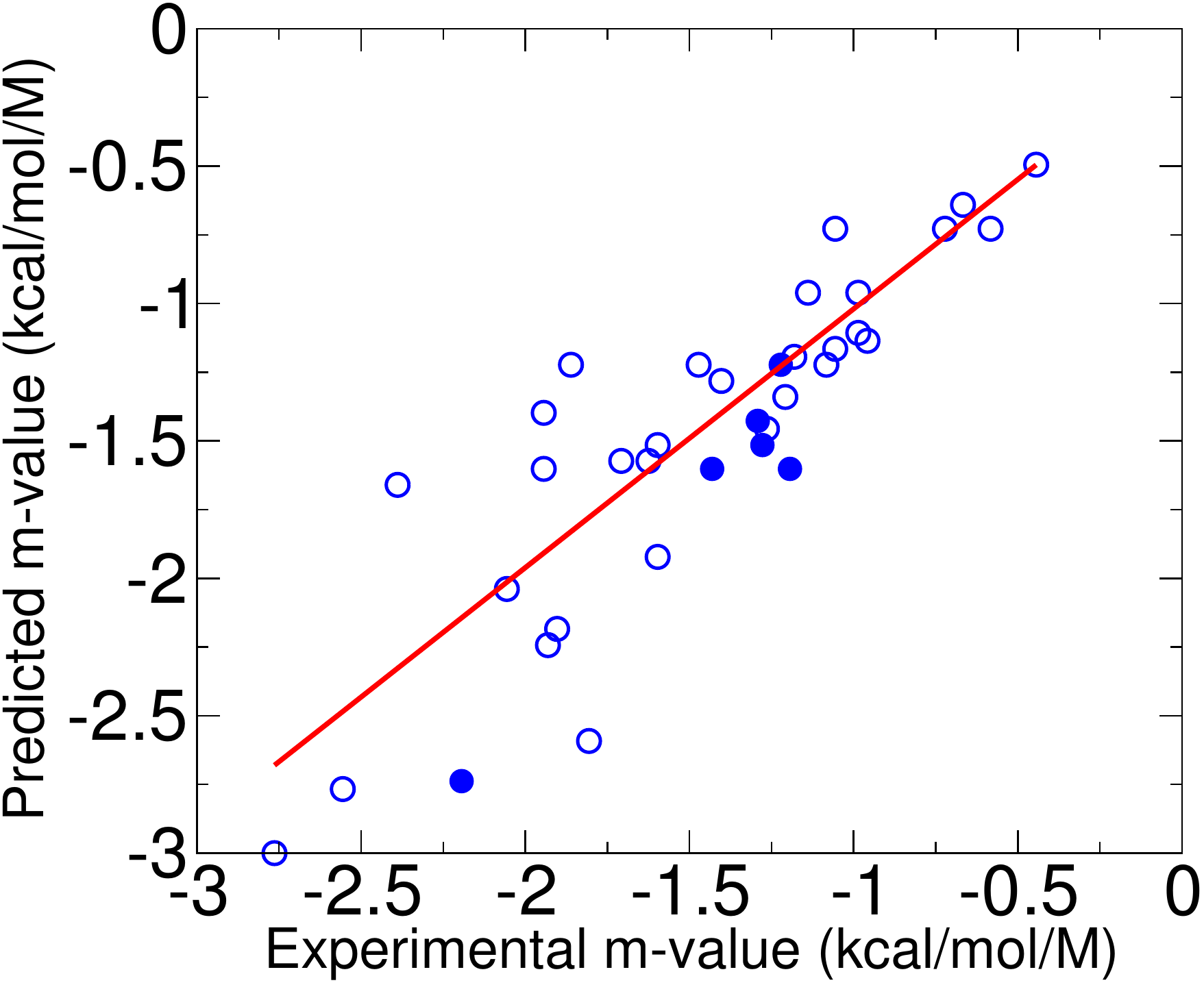}}\\
  \qquad \qquad
  \subfloat[]{\label{}\includegraphics[width=0.4\textwidth]{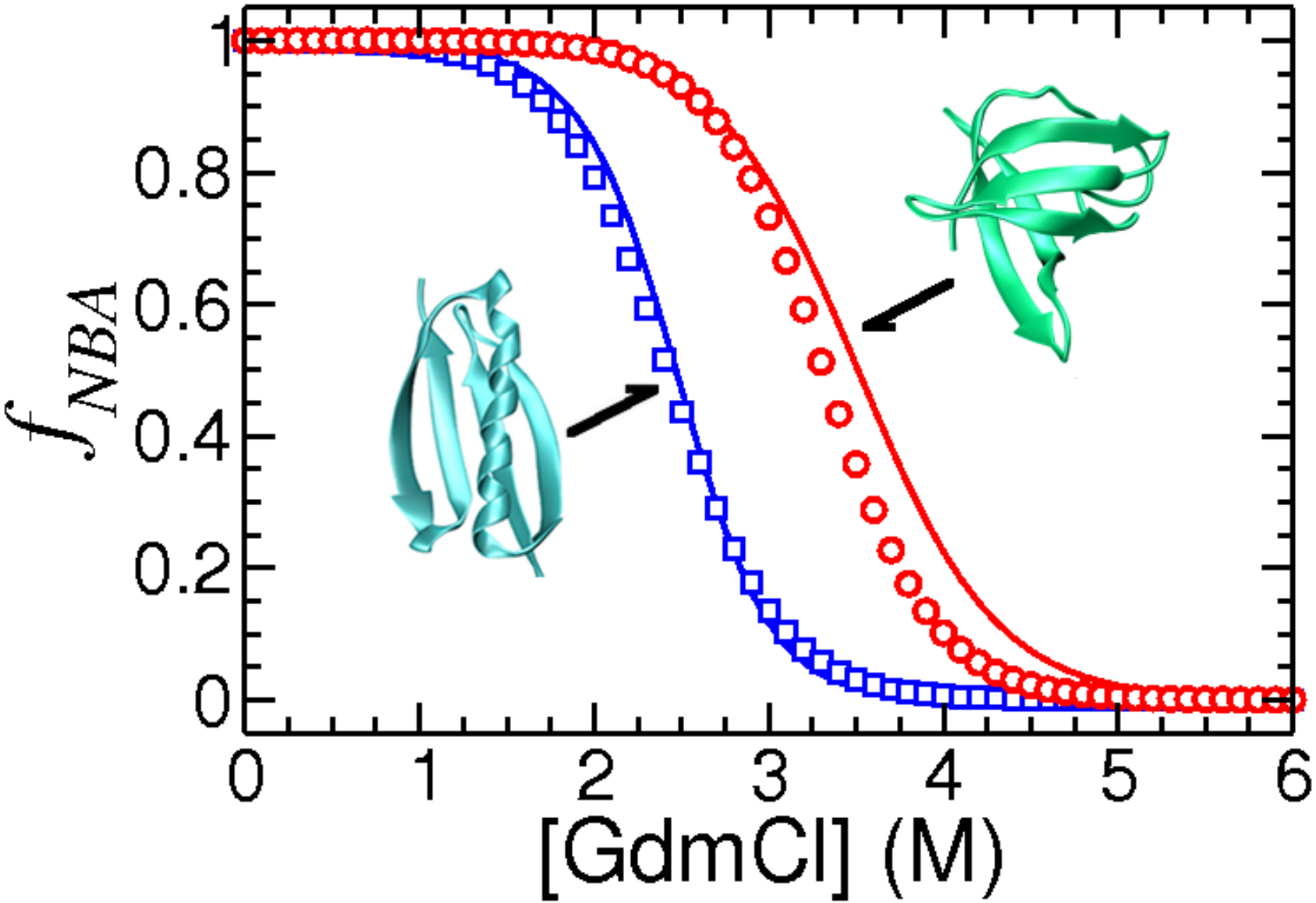}}&
  \qquad
  \subfloat[]{\label{}\includegraphics[width=0.4\textwidth]{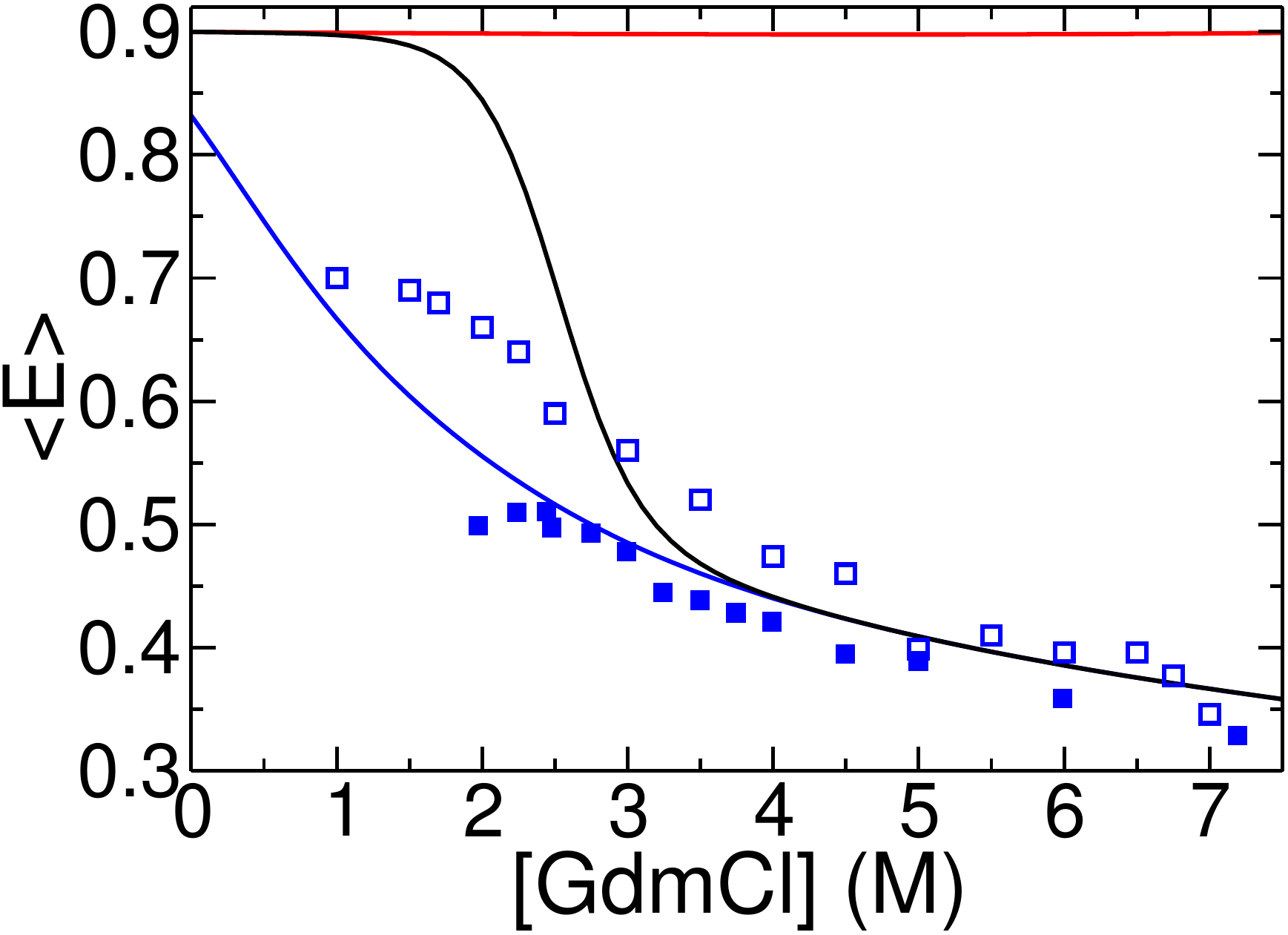}}
  \end{array}$
  \end{center}
 \caption{}
  \label{}
\end{figure}

\begin{figure}
  \begin{center}$
  \begin{array}{ccc}
    \subfloat[]{\label{}\includegraphics[width=0.38\textwidth]{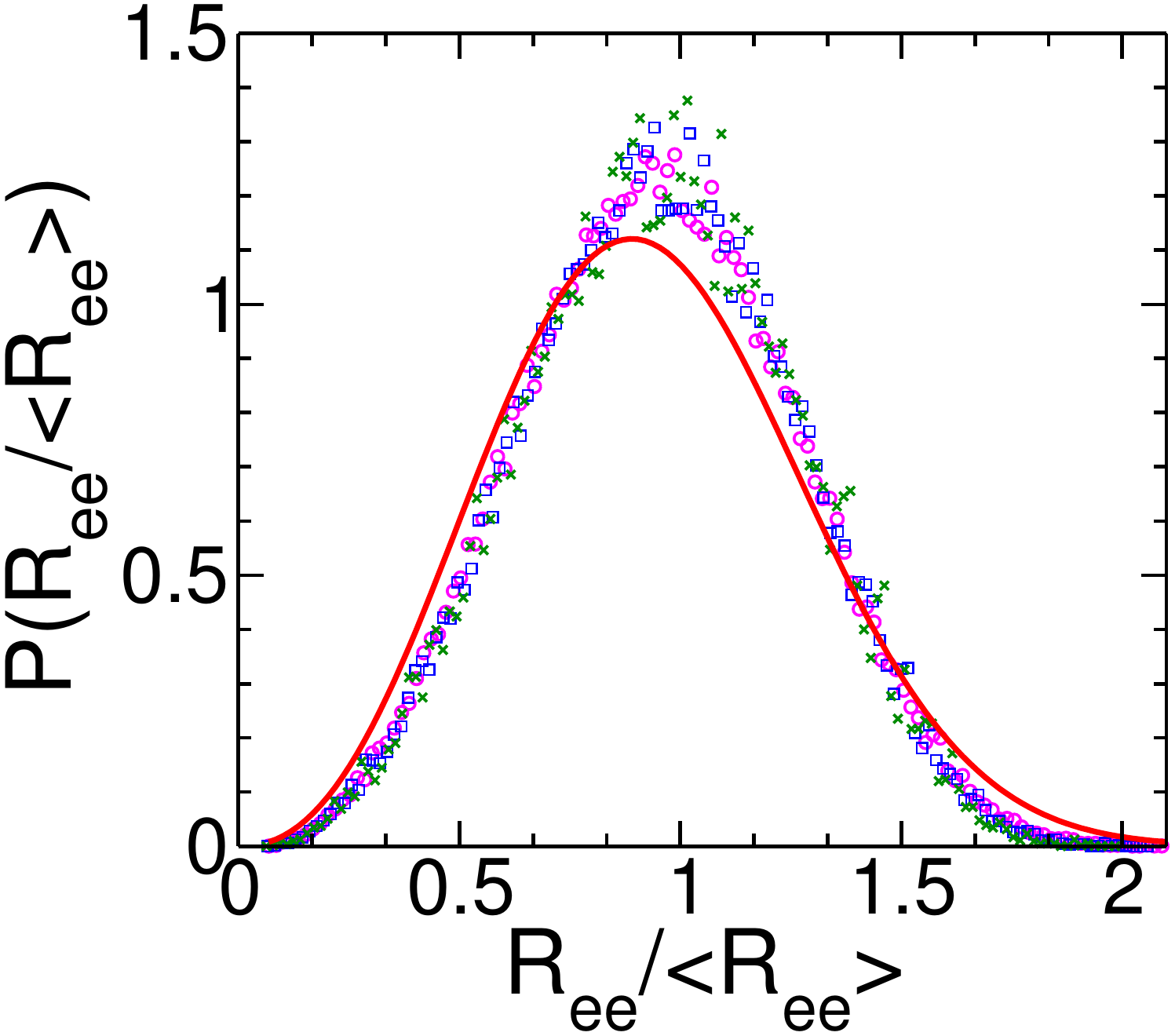}}&
    \subfloat[]{\label{}\includegraphics[width=0.42\textwidth]{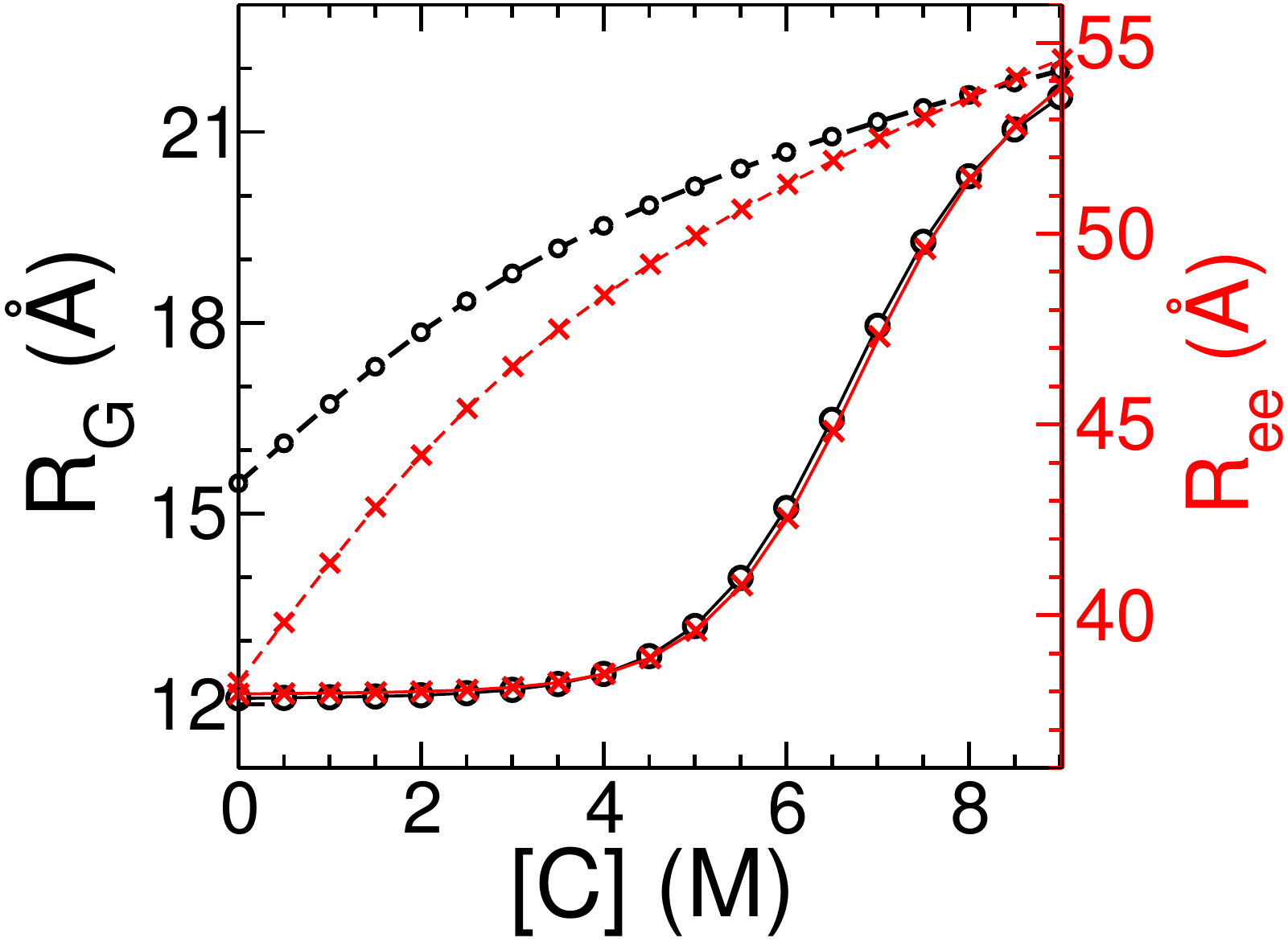}}&
    \subfloat[]{\label{}\includegraphics[width=0.4\textwidth]{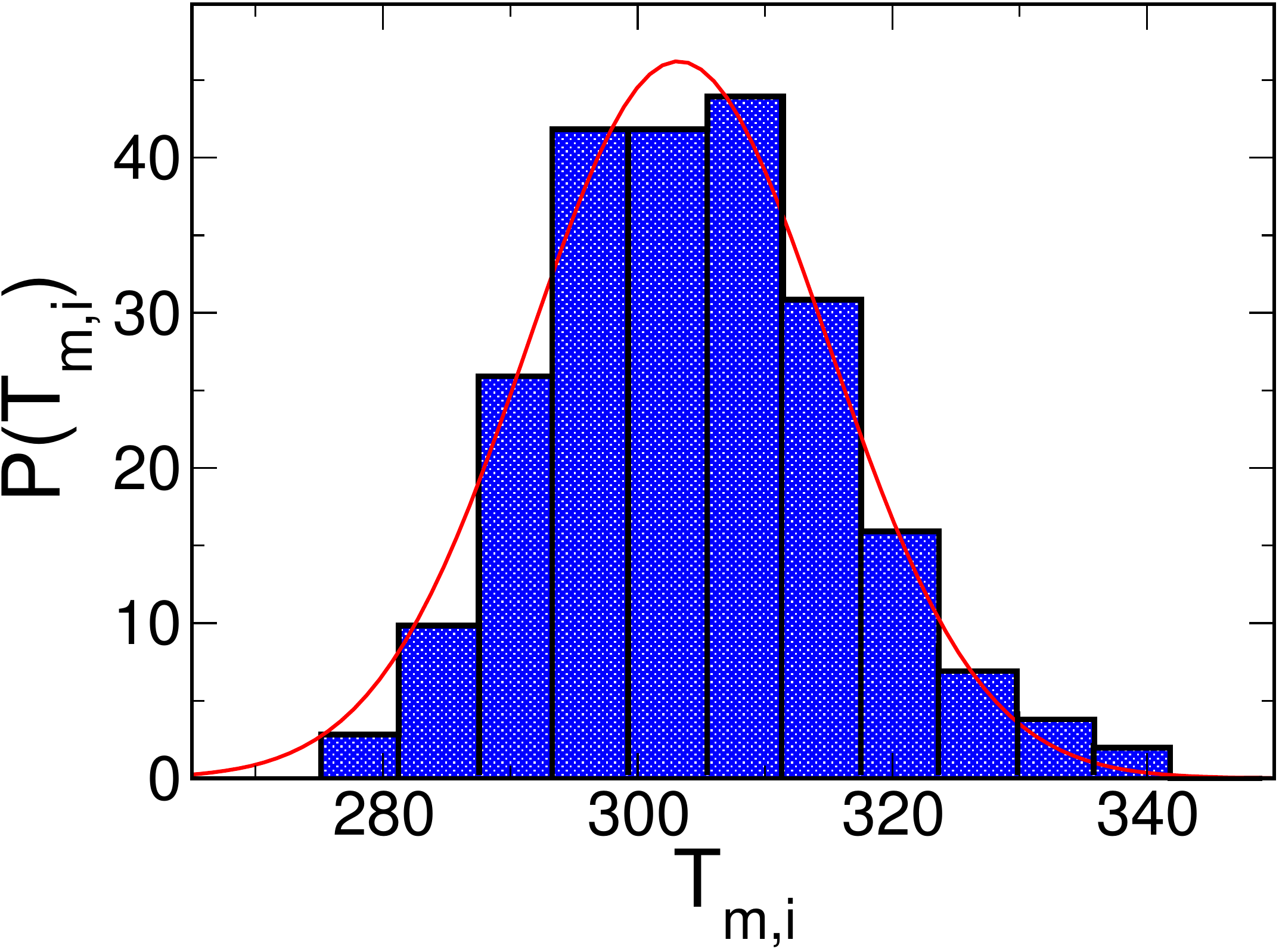}}\\
  \qquad \qquad
    \subfloat[]{\label{}\includegraphics[width=0.35\textwidth]{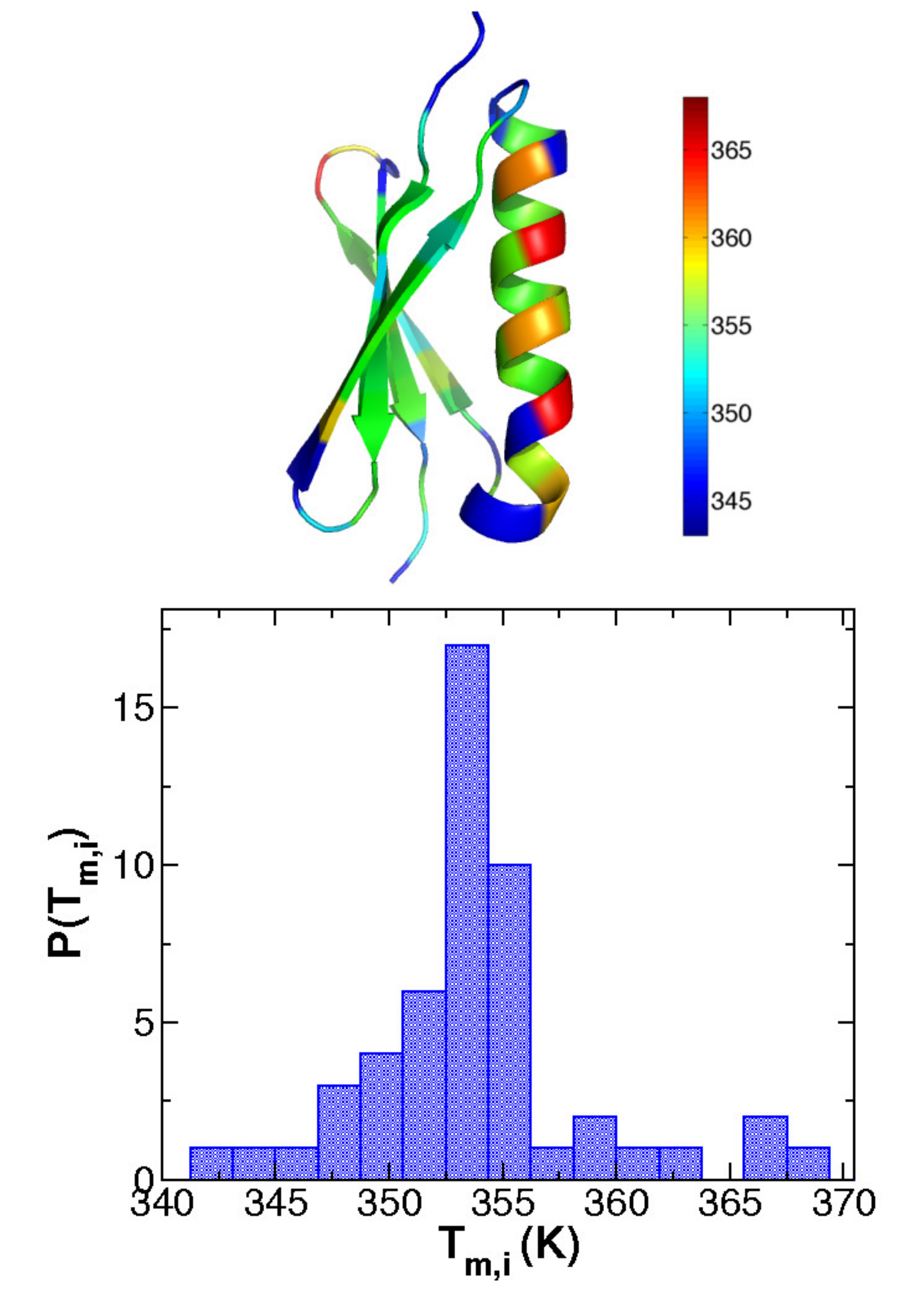}}&
    \qquad
  \subfloat[]{\label{}\includegraphics[width=0.35\textwidth]{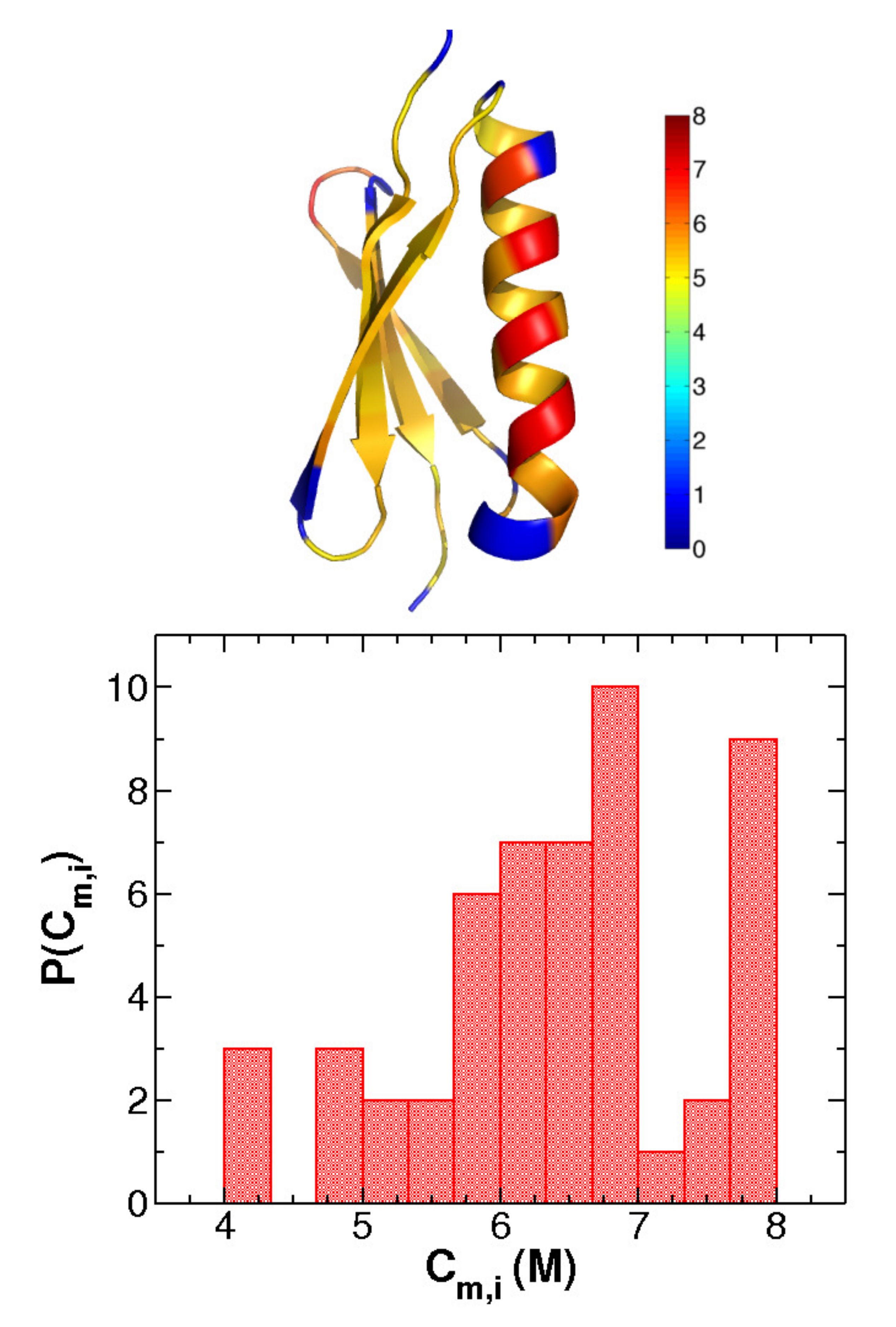}}&
    \subfloat[]{\includegraphics[width=0.4\textwidth]{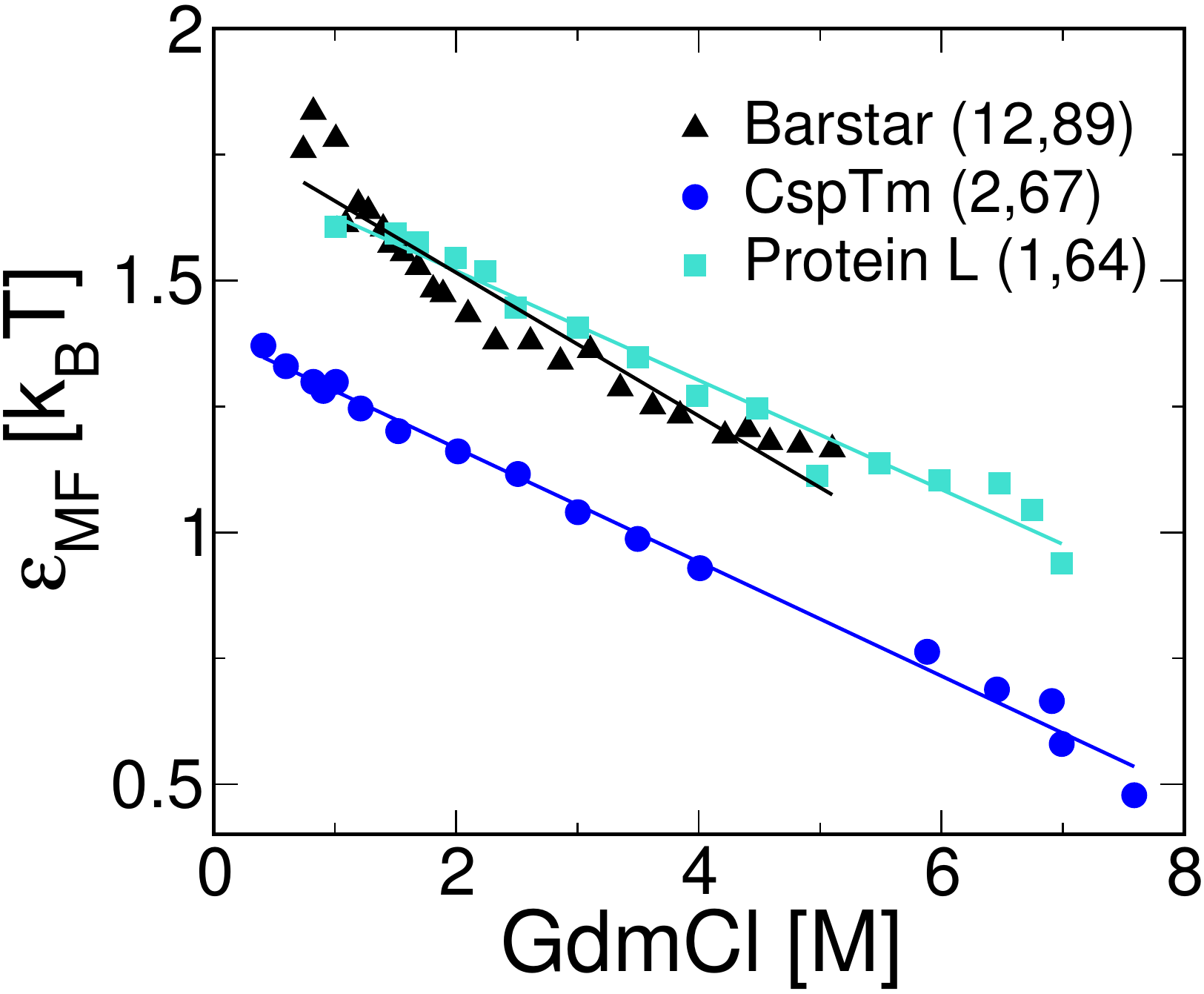}\label{}}
    \end{array}$
\end{center}
 \caption{}
  \label{}
\end{figure}

\begin{figure}
 \label{}\includegraphics[width=0.9\textwidth]{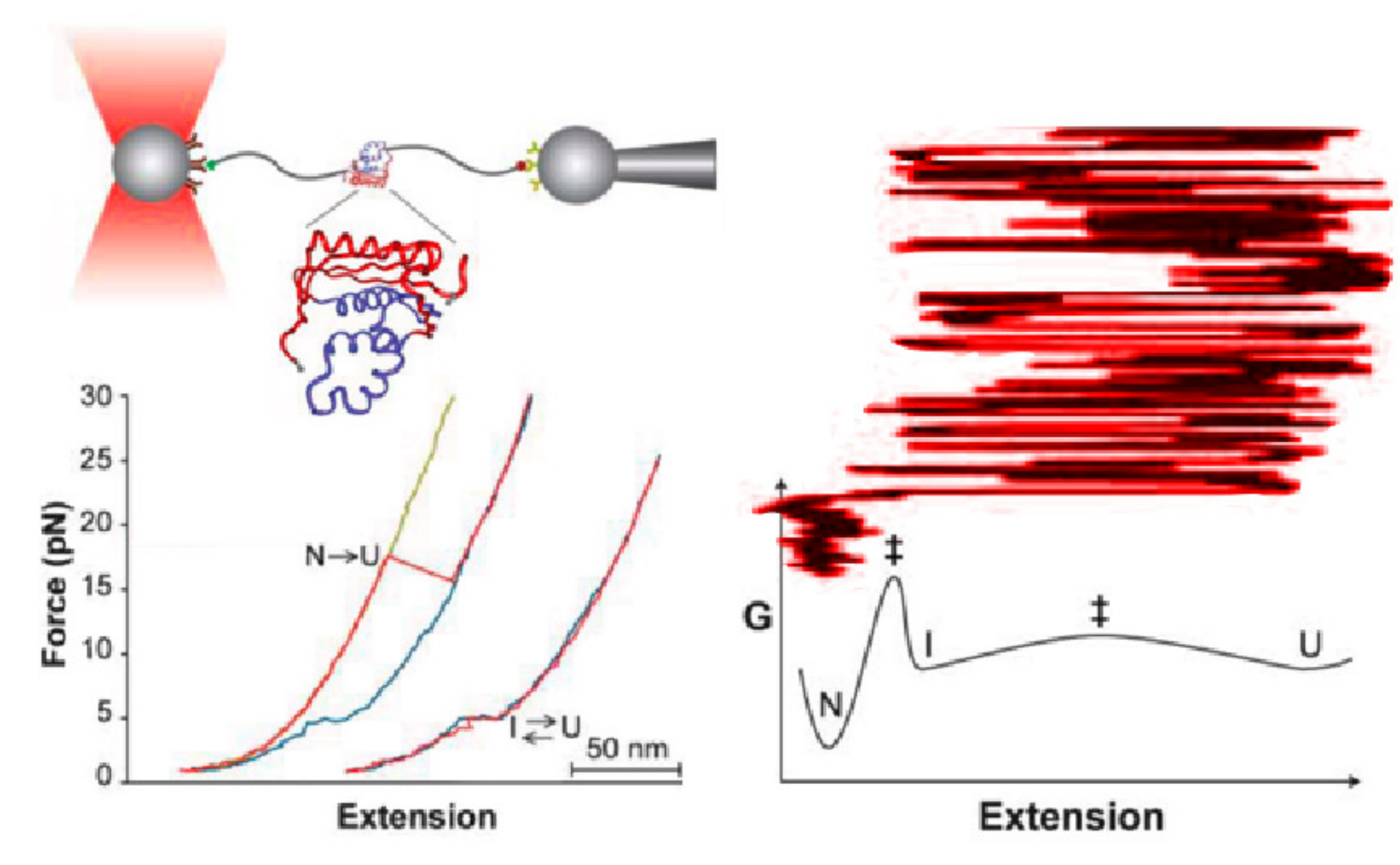}    
 \caption{}
  \label{}
\end{figure}

\begin{figure}
  \label{}\includegraphics[width=1.0\textwidth]{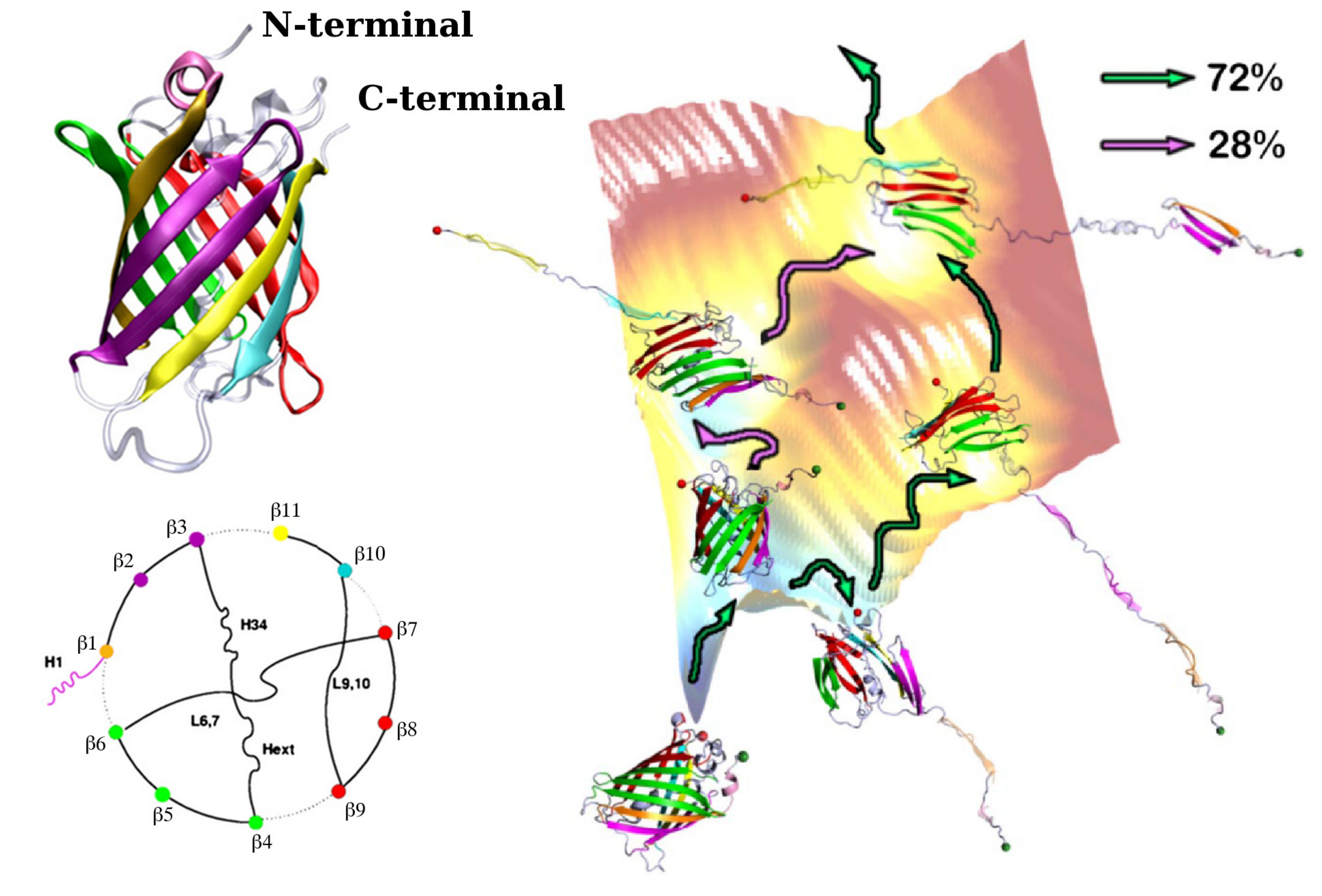}
  \caption{}
  \label{}
\end{figure}

\begin{figure}
\begin{center}$
  \begin{array}{c}
  \subfloat[]{\label{}\includegraphics[width=0.8\textwidth]{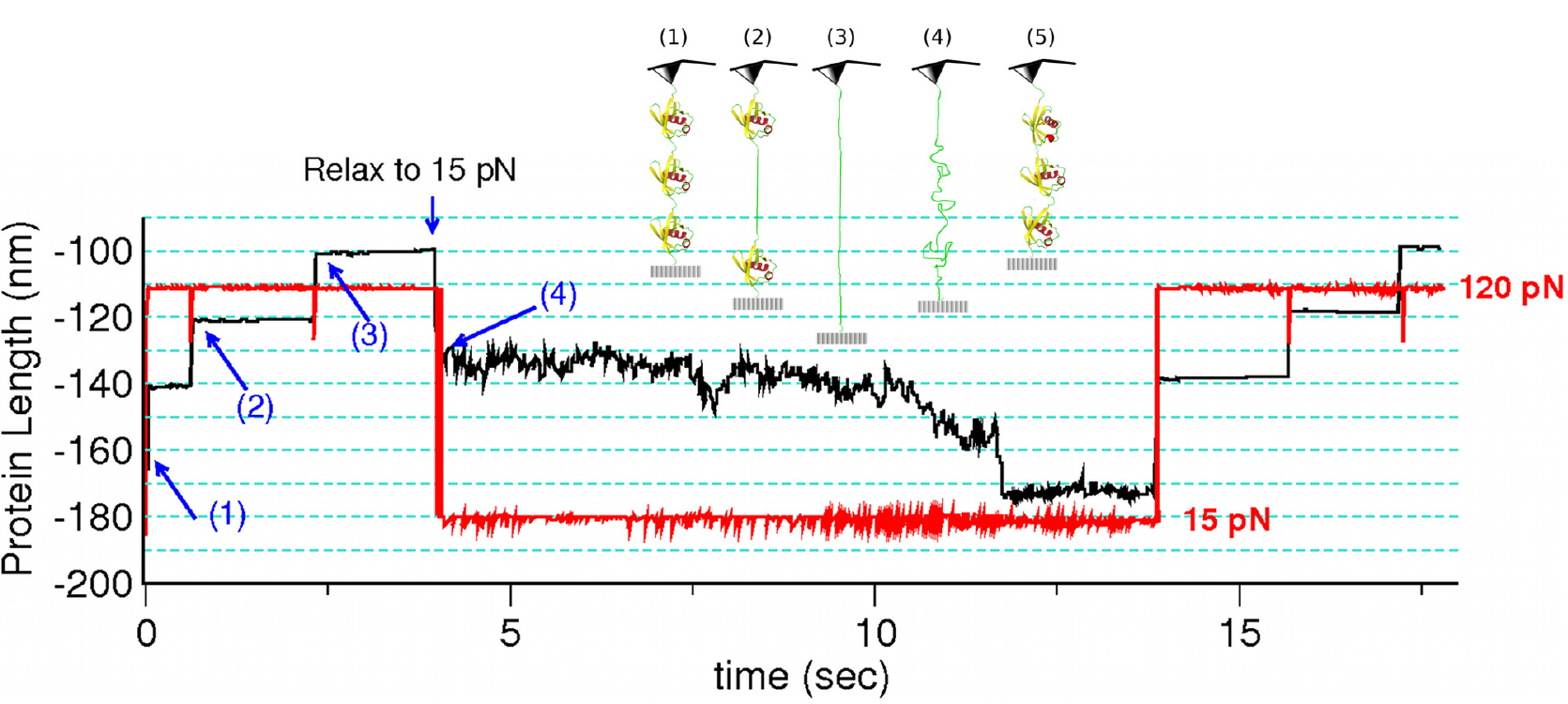}}\\
  \subfloat[]{\label{}\includegraphics[width=0.8\textwidth]{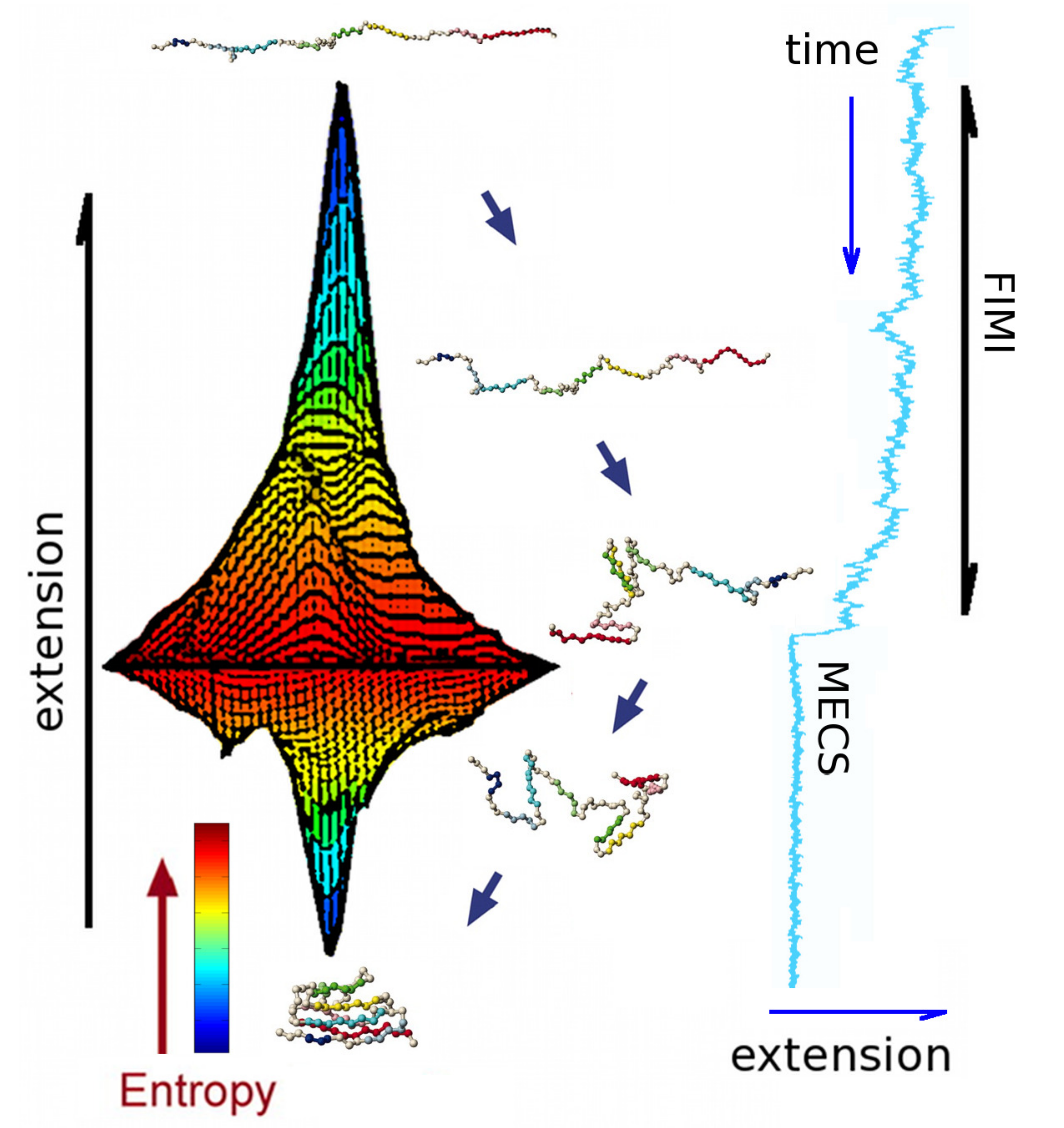}}
    \end{array}$
  \end{center}
  \caption{}
  \label{}
\end{figure}

\begin{figure}
\begin{center}$
  \begin{array}{c}
  \subfloat[]{\label{}\includegraphics[width=0.68\textwidth]{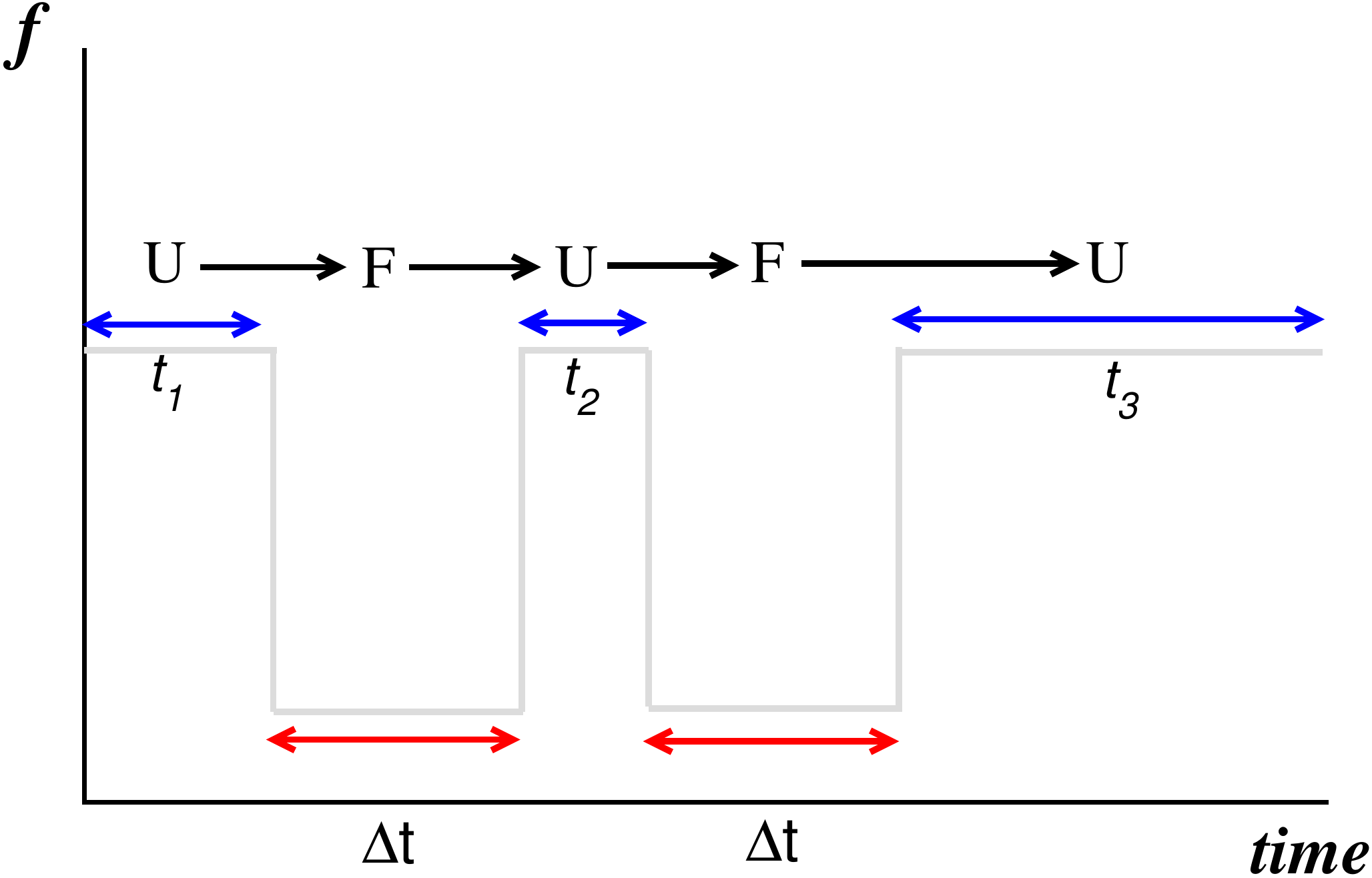}}\\
    \subfloat[]{\label{}\includegraphics[width=0.8\textwidth]{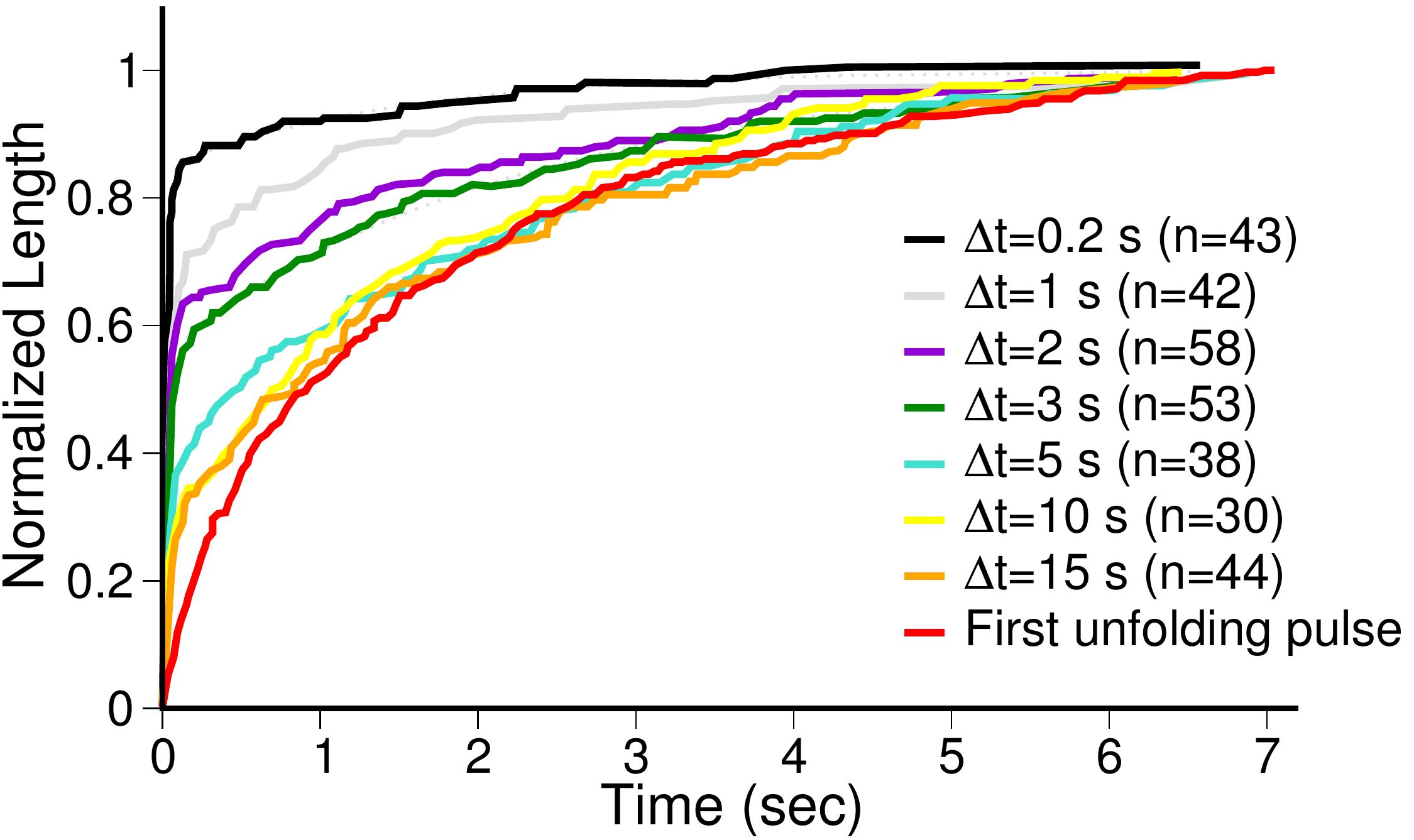}}  
    \end{array}$
  \end{center}
   \caption{}
  \label{}
\end{figure}

%


%


%
%
%
%
%
%
%
%
%
%
\end{document}